\documentclass{aa}  
\usepackage{graphicx}
\usepackage{txfonts}
\usepackage{amsmath}
\usepackage{amsfonts}
\usepackage{natbib}
\usepackage{longtable}
\usepackage[usenames, dvipsnames]{xcolor}
\usepackage[normalem]{ulem}

\begin{document}

   \title{IC~4665 DANCe}

   \subtitle{I. Members, empirical isochrones, magnitude distributions, present-day system mass function, and spatial distribution}

   \author{N. Miret-Roig\inst{1}
          \and H. Bouy \inst{1}
          \and J. Olivares\inst{1}
          \and L.~M. Sarro\inst{2}
          \and M. Tamura\inst{3,4,5}
          \and L. Allen\inst{6}          
		  \and E. Bertin\inst{7}
          \and S. Serre\inst{1}
          \and A. Berihuete\inst{8}
		  \and Y. Beletsky\inst{9}
          \and D. Barrado\inst{10}
          \and N. Hu\'elamo\inst{10}
          \and J.-C. Cuillandre\inst{11}
          \and E. Moraux\inst{12}          
          \and J. Bouvier\inst{12}
          }

   \institute{Laboratoire d'astrophysique de Bordeaux, Univ. Bordeaux, CNRS, B18N, allée Geoffroy Saint-Hilaire, 33615 Pessac, France.
         \email{nuria.miret-roig@u-bordeaux.fr}
         \and
         Depto. de Inteligencia Artificial , UNED, Juan del Rosal, 16, 28040 Madrid, Spain 
         \and
         Department of Astronomy, The University of Tokyo, 7-3-1, Hongo, Bunkyo-ku, Tokyo, 113-0033, Japan
         \and
         National Astronomical Observatory of Japan, 2-21-1, Osawa, Mitaka, Tokyo, 181-8588, Japan
         \and
         Astrobiology Center, 2-21-1, Osawa, Mitaka, Tokyo, 181-8588, Japan
         \and
         National Optical Astronomy Observatory, 950 North Cherry Avenue, Tucson, AZ 85719, USA
         \and
          Institut d'Astrophysique de Paris, CNRS UMR 7095 and UPMC, 98bis bd Arago, F-75014 Paris, France          
         \and
         Depto. Statistics and Operations Research, University of C\'adiz, Campus Universitario R\'\i o San Pedro s/n, 11510 Puerto Real, C\'adiz, Spain
         \and
         Las Campanas Observatory, Carnegie Institution of Washington, Colina el Pino, 601 Casilla, La Serena, Chile
         \and
         Centro de Astrobiología (CSIC-INTA), ESAC Campus, Camino Bajo del Castillo s/n, 28692 Villanueva de la Cañada, Madrid, Spain
         \and 
         AIM Paris Saclay, CNRS/INSU, CEA/Irfu, Universit\'e Paris Diderot, Orme des Merisiers, France 
         \and
         Univ. Grenoble Alpes, CNRS, IPAG, 38000 Grenoble, France
             }

   \date{Received ; accepted }

 
  \abstract
   {The study of star formation is extremely challenging due to the lack of complete and clean samples of young, nearby clusters, and star forming regions. The recent \textit{Gaia} DR2 catalogue complemented with the deep, ground based COSMIC DANCe catalogue offers a new database of unprecedented accuracy to revisit the membership of clusters and star forming regions. The 30~Myr open cluster IC~4665 is one of the few well-known clusters of this age and it is an excellent target where to test evolutionary models and study planetary formation.}
   {We aim to provide a comprehensive membership analysis of IC~4665 and to study the following properties: empirical isochrones, distance, magnitude distribution, present-day system mass function, and spatial distribution.}
   {We use the \textit{Gaia} DR2 catalogue together with the DANCe catalogue to look for members using a probabilistic model of the distribution of the observable quantities in both the cluster and background populations. } 
  {We obtain a final list of 819 candidate members which cover a 12.4 magnitude range ($7<J<19.4$). We find that 50\% are new candidates, and we estimate a conservative contamination rate of 20\%. This unique sample of members allows us to obtain a present-day system mass function in the range of 0.02--6~M$_\sun$, which reveals a number of details not seen in previous studies. In addition, they favour a spherically symmetric spatial distribution for this young open cluster.}
   {Our membership analysis represents a significant increase in the quantity and quality (low-contamination) with respect to previous studies. As such, it offers an excellent opportunity to revisit other fundamental parameters such as the age.}

   \keywords{Proper motions, Stars: luminosity function, mass function, Stars: low-mass, (Stars:) brown dwarfs, Galaxy: open clusters and associations: individual: IC~4665}

   \maketitle
%

\section{Introduction}

The Initial Mass Function (IMF), i.e. the frequency distribution of stellar masses at birth, is a fundamental parameter in the study of stellar formation and evolution. It was first introduced by \citet{Salpeter55} in the form of $\xi(\log_{10}m) = dn/d\log_{10}m$ and in this study we adopt the same formalism. Since then, strong efforts have been put on trying to constrain its shape in various environments. While it is fairly well known for intermediate masses, the extremes of the IMF remain uncertain. For the high mass domain the main difficulty is the low number of stars and their  fast evolution, while for the low mass domain the main difficulty is the high level of contamination and incompleteness even with the best photometric and astrometric surveys. 

The study of the IMF requires an accurate and comprehensive census of the cluster members. In this study, we propose to derive such a census of one nearby young cluster namely, IC~4665. For this purpose, we use wide-field deep catalogues encompassing a large area around the cluster, and analyse them using a powerful classification algorithm capable of identifying the few hundreds of cluster members within the millions of interlopers. Hereafter, we refer as members to the candidate members resultant from the membership analysis. 

There are few well-known, nearby ($<500$~pc), pre-main sequence open clusters in the age interval 10--50~Myr. One of them is IC~4665, located in the Ophiuchus constellation and first reported by Philippe Loys de Chéseaux in 1745. Its age was estimated using the lithium depletion boundary at $27.7^{+4.2}_{-3.5}$ Myr by \cite{Manzi+08}. From pre-main sequence isochrone fitting and upper-main-sequence turn-off fitting, \citet{Cargile+10} derived an age and a distance of $36\pm 9$~Myr and $360\pm 12$~pc and $42\pm12$~Myr and $357\pm12$~pc, respectively. 

The first study of the IMF of IC~4665 was carried out by \citet{deWit+06}. They selected their members from photometric observations in the optical obtained at the Canada-France-Hawaii Telescope (CFHT). They estimated a contamination by foreground and background stars of up to 85\% using control fields, which can be explained by its low galactic latitude. They reported a mass function best described by a power law with an exponent of --0.6 for the low mass objects down to $\sim$0.1~M$_\sun$. Later, \citet{Lodieu+11} performed a similar analysis adding near-infrared photometry from the UKIRT Infrared Deep Sky Survey \citep[UKIDSS,][]{Lawrence+07} to the previous observations of \citet{deWit+06}. They revised the members of previous studies as well as proposed new candidate members. They reported a mass function best represented by a log-normal function with a peak at 0.25--0.16~M$_\sun$. The differences between the mass functions obtained with these two studies can be mainly attributed to the large contamination rate of field stars, as we shall see later.

Recently, the second \textit{Gaia} data release (\citealt{GaiaColBrown+18}, hereafter \textit{Gaia} DR2) was public providing the five-parameter astrometric solution (positions on the sky, parallaxes, and proper motions) as well as $G$, $G_{\rm BP}$ and $G_{\rm RP}$ magnitudes for more than 1.3 billion sources, with a limiting magnitude of $G\approx21$~mag at the faint end and $G\sim 3$~mag at the bright end. The average astrometric precision is of the order of the mas~yr$^{-1}$ in proper motion and below the mas in parallax, and the average photometric precision is at the milli-magnitude level. This constitutes an unprecedented accurate astrometric+photometric dataset ideally suited to study the census of nearby open clusters. A first demonstration of the power of the \textit{Gaia} data is presented in \citet{GaiaColBabusiaux+18}. They studied the fine structures of the Hertzsprung-Russell diagram (HRD) in the field and in open and globular clusters. IC~4665 was among their targets and they provided a list of 174 high-probability members up to magnitude $G<18$. Soon after, \citet{Cantat-Gaudin+18} presented another study of open clusters using \textit{Gaia} DR2 data. They derived another membership list (with the same magnitude limit) made of 175 high-probability members, 146 of which are in common with \citet{GaiaColBabusiaux+18}. Both studies used only the \textit{Gaia} data, applied a strict filtering, and discarded sources fainter than $G=18$~mag, thus delivering a clean but yet highly incomplete sample. 

Over the past few years, \citet{Bouy+13} started a survey program, the DANCe project, with the aim of deriving a comprehensive and homogeneous census of stellar and sub-stellar sources in the nearby ($<1$~kpc), young ($<500$~Myr) clusters. This survey combines deep, wide-field, multi-epoch images obtained at various observatories to build a catalogue of accurate proper motions and multi-wavelength photometry with a sensitivity up to 5 magnitudes deeper than \textit{Gaia} and including near-infrared data. Here we present the DANCe catalogue for the region of IC~4665. After identifying candidate members, we study the cluster properties and in particular its empirical isochrones, distance, magnitude distribution, present-day mass function (PDMF), and spatial distribution.
The assumptions in this work are based on the properties of the dataset and the cluster. We strongly recommend the reader to look at Sect.~2 of \citet{Olivares+19} for more details.

This paper is structured as follows. First, we introduce the two datasets used in this study (Sect.~\ref{sec:data}). Second, we present the algorithm we use for the membership analysis (Sect.~\ref{sec:membership}) and discuss the results obtained including a comparison with previous studies (Sect.~\ref{sec:analysis-members}). Then, we provide the empirical iscohrones of this young cluster and compare them with evolutionary models (Sect.~\ref{sec:isochrones}). In addition, we compute the apparent magnitude distribution and the PDMF (Sect.~\ref{sec:mass-func}). Finally, we study the spatial distribution of the cluster (Sect.~\ref{sec:spatial_dist}), and we draw our conclusions (Sect.~\ref{sec:conclusions}).


\section{The data}
\label{sec:data}
In this work we used two different datasets with different origins and properties to look for members in the IC~4665 open cluster. In this section, we describe how we obtained each of them. 

\subsection{The Gaia DR2 dataset \label{sec:gaiadr2}}
We queried a circular area of 3\degr\, radius around the centre of the cluster (RA = 266.6\degr , Dec = 5.7\degr), from the \textit{Gaia} DR2 catalogue (see Appx.~\ref{app:query_GDR2}). Then, we only kept the sources with a full five-parameter solution available. Some quality checks have been suggested in the literature. We did not apply any filtering techniques because we want to be as complete as possible. The filtering recommended by the \textit{Gaia} team is based on the renormalised unit weighted error (RUWE) and is described in detail in a publicly available technical note\footnote{\url{https://www.cosmos.esa.int/web/gaia/dr2-known-issues}}. The RUWE criterion is a quality indicator which might be used when the aim is to have only the most precise, reliable, and consistent astrometric solutions. However, it also leads to a higher degree of incompleteness. For instance, since the \textit{Gaia} DR2 catalogue does not deal with binaries, their solution is likely to be "inconsistent", and thus the RUWE filter will remove most of the already few binaries included in \textit{Gaia} DR2. We therefore have no strong scientific argument to cut our sample by this or any other kind of filtering. In addition, the sources with problematic astrometric solutions, may be rejected later on based on complementary observations and/or subsequent \textit{Gaia} DRs.

This sample contains positions, proper motions, parallaxes, and  $G,G_{\rm BP},G_{\rm RP}$ photometry for 1\,217\,725 sources and, hereafter, we refer to it as the GDR2 catalogue\footnote{ By GDR2 we refer to the IC~4665 catalogue based on \textit{Gaia} DR2 data. Not to be confused with the full ESA catolgue to which we refer as the \textit{Gaia} DR2 catalogue.}. The mean errors of this catalogue are $\sim0.5$ mas for parallaxes, $\sim1$ mas yr$^{-1}$ for proper motions, and $<0.1$ mag for the photometry. According to \citet{GaiaColBrown+18}, the catalogue is mostly complete down to  $G=7$~mag. On the faint side, \cite{Lindegren+18} reported that the five-parameter solution is 94.5\% complete up to $G=19$~mag (see their Table~B.1). In the following, we therefore assume that the GDR2 catalogue is complete between $7<G<19$~mag.

\subsection{The DANCe dataset}
\label{subsec:DANCe_dataset}

\begin{table*}
\caption{Instruments used in this study. }
\setlength\tabcolsep{4.5pt}
\begin{tabular}{lcccccc}\hline\hline
Telescope   & Instrument        & Filters          & Platescale     & Field of view & Epoch Min./Max.  & Ref. \\
            &                   &                  & [pixel$^{-1}$] &               &                  &      \\
\hline
CTIO (Blanco)  & Flaugher+10   & $g,r,i,z,y$  & 0\farcs27 & 1.1\degr~radius & 2014--2018  & (1) \\
KPNO (Mayall)  & NEWFIRM   & $J,H,Ks$  & 0\farcs4 & 28\arcmin$\times$28\arcmin & 2015  & (2) \\
CFHT   & MegaCam   & $r,i$  & 0\farcs18 & 1\degr$\times$1\degr & 2005--2015  & (3) \\
CFHT   & WIRCam   & $y,J,H,Ks$  & 0\farcs3 & 20\arcmin$\times$20\arcmin & 2007--2008  & (4) \\
CFHT   & CFH12K   & $I,z$  & 0\farcs21 & 42\arcmin$\times$28\arcmin & 1999--2002  & (5) \\
INT & WFC & $u,v,b,\beta,y$ (Str\"omgren), $U,B,V,Z,g,r,i$\tablefootmark{a}   & 0\farcs33 & 34\arcmin$\times$34\arcmin & 2000--2015 & (6) \\
UKIRT &  WFCAM & $J,Ks$ & 0\farcs4 & 40\arcmin$\times$40\arcmin \tablefootmark{b} & 2006-2012 & (7)\\
LCO Swope & Direct CCD & $i$ & 0\farcs43 & 15\arcmin$\times$14\arcmin  & 2013 & (8) \\
VST & OmegaCam & $r$ & 0\farcs21 & 1\degr$\times$1\degr & 2014 & (9) \\
ESO (2.2m) & WFI & $R,I$~\tablefootmark{a} & 0\farcs24 & 34\arcmin$\times$33\arcmin & 2002 & (10) \\
Subaru & HSC & $y$ & 0\farcs17 & 1.8\degr~radius & 2015 & (11) \\
Palomar 48" & PTF & $g,r$ & 1\farcs0 & 3\fdg3$\times$2\fdg2\tablefootmark{c} & 2010--2012 & (12) \\
OMM (1.6m) & CPAPIR & $I,J,H$ & 0\farcs89 & 30\arcmin$\times$30\arcmin & 2012--2015 & (13) \\
\hline
\hline
\end{tabular}

\noindent\tablefoottext{a}{as well as various narrow and medium bands}\\
\tablefoottext{b}{the chip layout has large gaps between detectors, and the coverage of the focal plane is only partial}\\
\tablefoottext{c}{one of the 12 detectors is dead}
\tablebib{(1)~\citet{Flaugher+10}; (2)~\citet{Autry+03}; (3)~\citet{Boulade+03}; (4)~\citet{Thibault+03}; (5)~\citet{Cuillandre+00}; (6)~\citet{Ives98}; (7)~\citet{Casali+07}; (8)~\citet{Rheault+14}; (10)~\citet{Baade+99}; (11)~\citet{Miyazaki+18}; (12)~\citet{Rahmer+08}; (13)~\citet{Thibault+02}}
\label{tab:observations}
\end{table*}

\begin{figure}
\begin{center}
\includegraphics[width = \columnwidth]{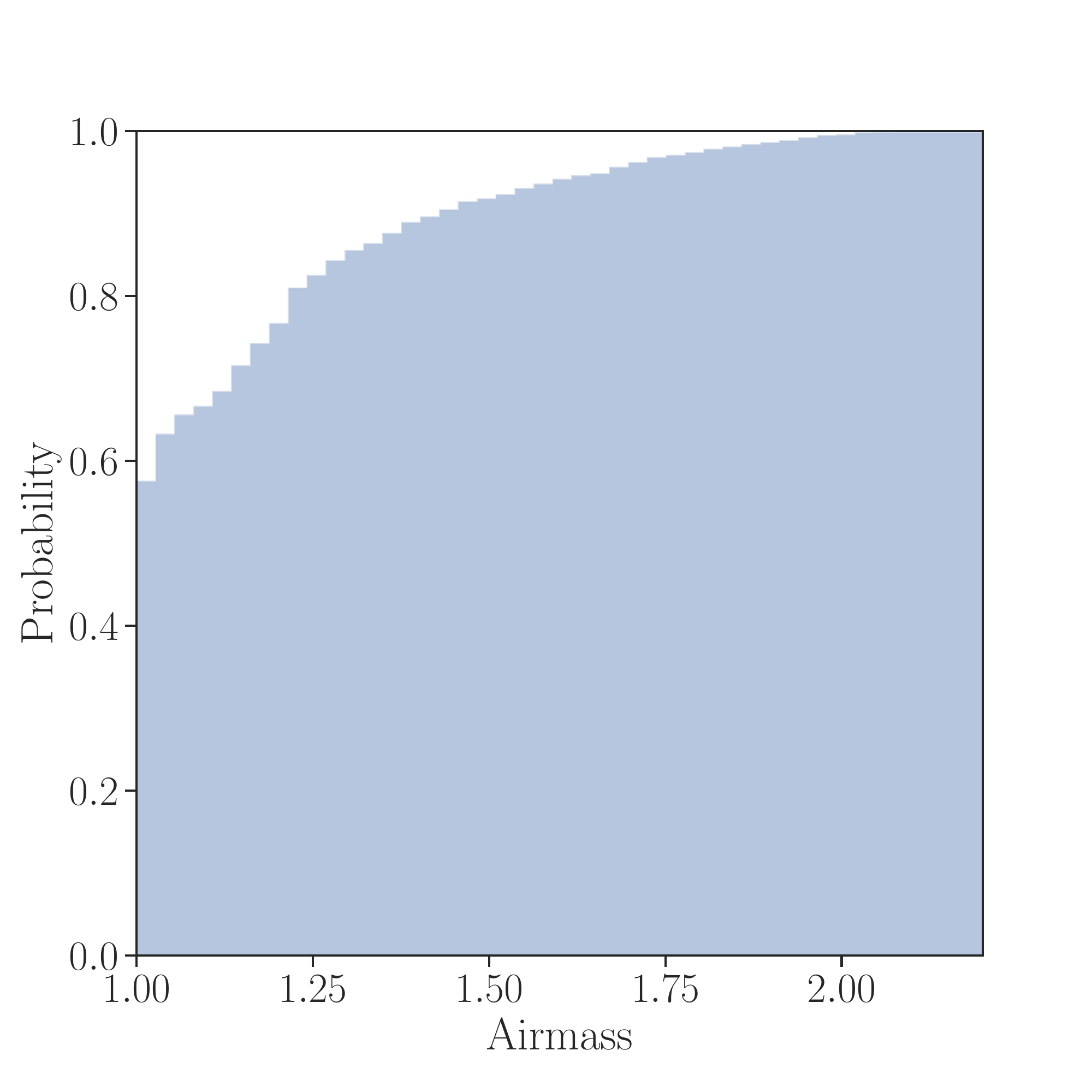}
\includegraphics[width = \columnwidth]{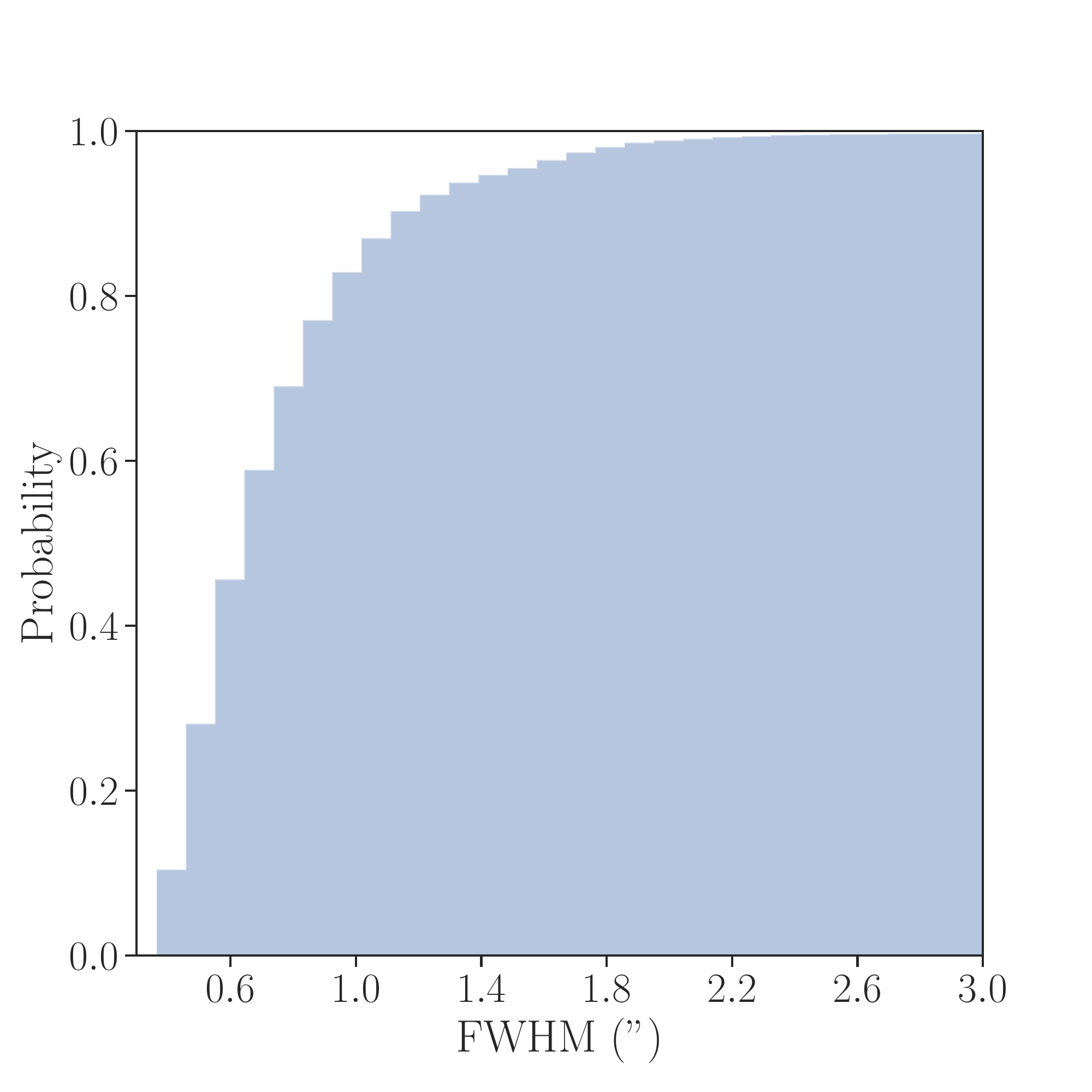}
\caption{Top: Cumulative distribution of airmass for the observations. Bottom: Cumulative distribution of average FWHM for the images. }
\label{fig:airmass-fwhm}
\end{center}
\end{figure}

\begin{figure}
\begin{center}
\includegraphics[width = \columnwidth]{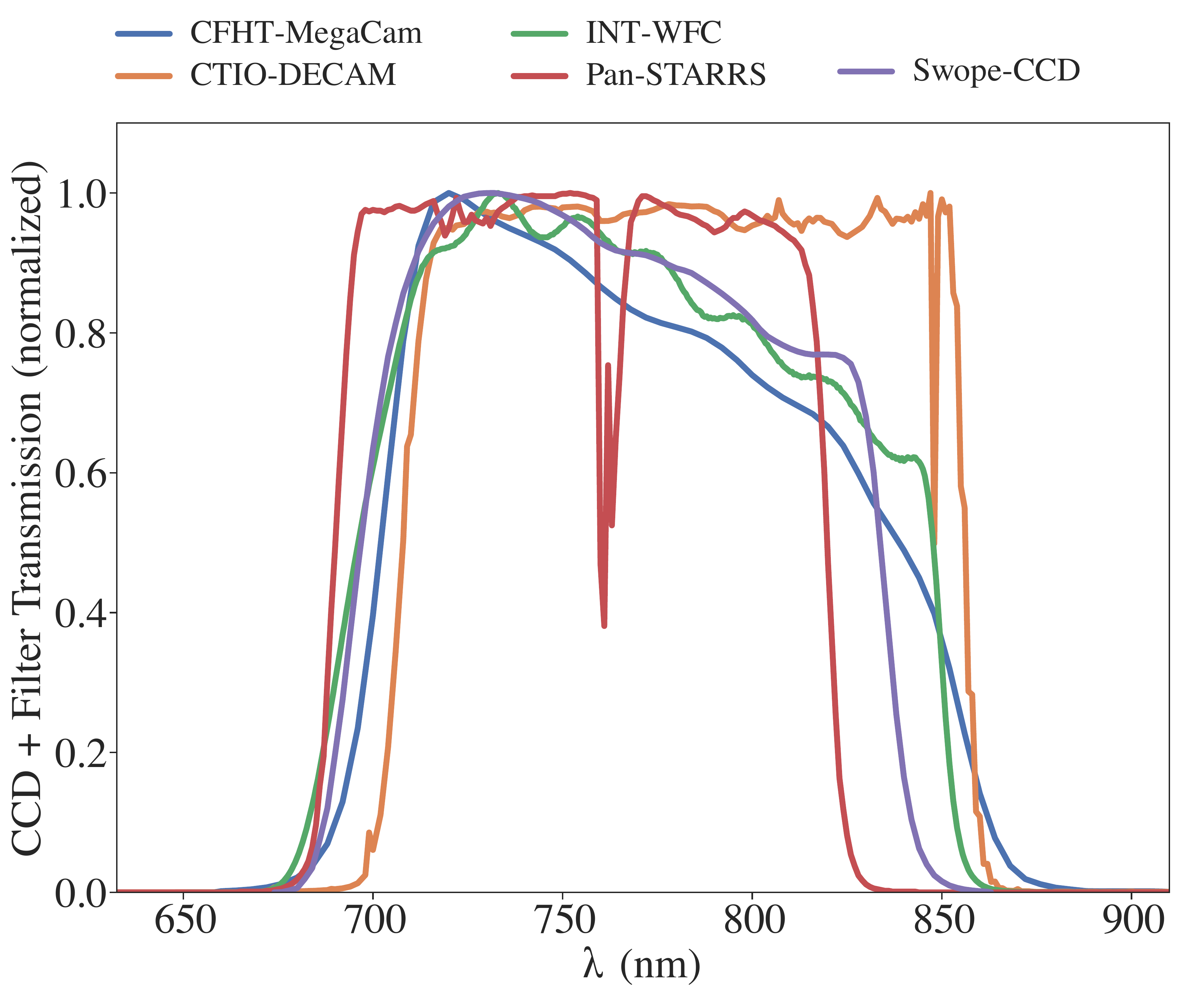}
\caption{Transmission curves of the $i$ filters used in this study.}
\label{fig:iband}
\end{center}
\end{figure}

We searched the European Southern Observatory (ESO) archive, the National Optical Astronomy Observatory (NOAO) archive, the PTF archive hosted at the NASA/IPAC Infrared Science Archive (IRSA), the Canadian Astronomy Data Centre (CADC) archive, the Isaac Newton Group (ING) archive, the WFCAM Science (WSA) archive, and the SMOKA science archive for wide field images within a circular region of 3\degr\ radius, centred on IC~4665. The data found in these public archives was complemented with our own observations using the Las Campanas Swope telescope and its Direct CCD camera, the Dark Energy Camera (DECam) mounted on the Blanco telescope at the Cerro Tololo Inter-American Observatory (CTIO), the NEWFIRM camera mounted on the 4m telescope at the Kitt Peak National Observatory (KPNO), the Hyper Suprime-Cam (HSC) mounted on the Subaru telescope at the National Astronomical Observatory of Japan, and the Wide Field Camera (WFC) mounted on the Isaac Newton Telescope (INT). A number of observations found in the archives were discarded after a visual inspection, because of their poor quality, limited sensitivity or acquisition problems. Table~\ref{tab:observations} gives an overview of the various cameras used for this study. 

The airmass and the Full-Width at Half Maximum (FWHM) measured in the images using point-like sources are two important parameters influencing the achievable astrometric accuracy.
About 90\% of the observations were obtained at airmass $\le 1.5$ (see Fig.~\ref{fig:airmass-fwhm} top). IC~4665 is located at a declination of $\delta\sim$~5\degr and we gathered data from both hemispheres. 
About 82\% of the images have FWHM $\le$ 1\arcsec, and 90\% have FWHM $\le$ 1\farcs2 (see Fig.~\ref{fig:airmass-fwhm} bottom). 

In all cases except for MegaCam, WIRCam, PTF, DECam, UKIRT and HSC images, the raw data and associated calibration frames were downloaded and processed using standard procedures using an updated version of \emph{Alambic} \citep{Vandame02}, a software suite developed and optimised for the processing of large multi-CCD images. In the case of CFHT/MegaCam, the images processed and calibrated with the \emph{Elixir} pipeline were retrieved from the CADC archive \citep{Magnier+04}. The WIRCam images processed with the official \textit{'I'iwi} pipeline were retrieved from the CADC archive. In the case of DECam, the images processed with the community pipeline \citep{Valdes+14} were retrieved from the NOAO public archive. The pipeline processed PTF images were downloaded from the IPAC archive. UKIRT images from the UKIDSS and UHS surveys \citep{Dye+18} processed by the Cambridge Astronomical Survey Unit were retrieved from the WFCAM Science Archive. Finally, the HSC raw images were processed using the official HSC pipeline \citep{Bosch+18}.

\subsubsection{Astrometric analysis}
\label{subsec:astrometric_analysis}

After a visual rejection of problematic images (mostly due to loss of guiding, tracking or electronics problems), the dataset included 6\,774 individual images originating from 13 instruments. The total amount of data (scientific images, associated calibrations, and intermediate products) added to almost 20TB. 

The astrometric calibration was performed as described in \citet{Bouy+13}. The recently released \textit{Gaia} DR2 catalogue was used as external astrometric reference instead of the 2MASS catalogue, leading to a much improved astrometric solution. 

The final average internal and external 3$\sigma$ residuals  add up to $\sim$25~mas for high signal-to-noise (photon noise limited) sources. As explained in \citet{Bouy+13}, the proper motions computed are relative and display an offset with respect to the ICRS. We estimate the offset by computing the median offset between our and the \textit{Gaia} DR2 proper motion measurements after rejecting outliers using the modified Z-score \citep{Iglewicz+93}. We find offsets of $(\Delta\mu_{\alpha}\cos\delta,\Delta\mu_{\delta})=(1.70,4.48)$~mas~yr$^{-1}$. The uncertainty on this offset is estimated using bootstrapping and is found to be negligible ($<$0.003~mas~yr$^{-1}$).

Given the superiority and robustness of \textit{Gaia} measurements compared to our ground-based measurements, the \textit{Gaia} DR2 proper motion measurements are always preferred when available. Therefore, we cross-matched our catalogue with the \textit{Gaia} DR2 using a 1\arcsec\, radius and we kept all the \textit{Gaia} proper motions when available. The median uncertainties in proper motions are $\sim$2~mas~yr$^{-1}$. 

We found that about 1.3\% of the sources ($\sim$60\,000) were duplicated in the final catalogue. A visual inspection showed that they were almost all very low signal-to-noise and that the \textsc{SExtractor} deblending algorithm resolved them as two sources instead of one, in one (or a few) images. These resolved sources later fooled the cross-identification algorithm and ultimately resulted in two independent sources instead of one. There is no straightforward solution to this problem as for now, but given their very small number we treated them as regular sources in the rest of the analysis and simply looked for duplicated entries in the final members list.

\subsubsection{Photometric analysis}

\begin{figure*}
\begin{center}
\includegraphics[width = 1\textwidth]{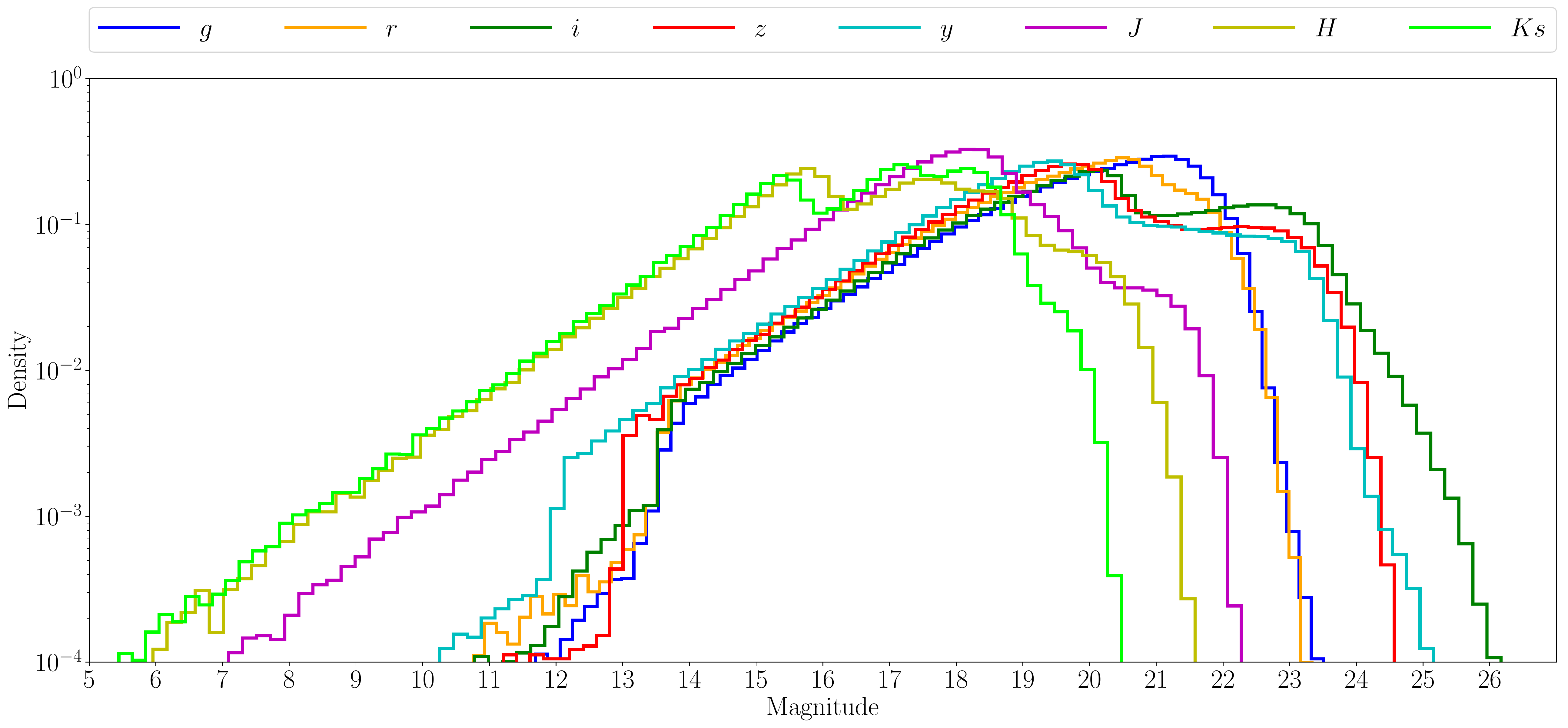}
\caption{Density of sources as function of the magnitude for the DANCe catalogue.}
\label{fig:completeness}
\end{center}
\end{figure*}

\begin{table*}[ht]
\centering
\caption{Number of measurements and percentage for each photometric band in the DANCe catalogue. The total number of sources is 2\,358\,937. }
\label{table:thresholds}
\begin{tabular}{l|rrrrrrrr}
  \hline
  \hline  
 &
\multicolumn{1}{c}{$g$} & \multicolumn{1}{c}{$r$} & \multicolumn{1}{c}{$i$} & \multicolumn{1}{c}{$z$} & \multicolumn{1}{c}{$y$} & \multicolumn{1}{c}{$J$} & \multicolumn{1}{c}{$H$} & \multicolumn{1}{c}{$Ks$} \\[0.5ex] 
 \hline
 \hline 
Num. obs.  & 1\,284\,683 & 1\,570\,253 & 2\,295\,949 & 2\,096\,267 & 2\,019\,385 & 1\,717\,645 & 888\,020 & 869\,769  \\
Percentage & 54\%        & 67\% & 97\% & 89\% & 86\% & 73\% & 38\% & 37\%  \\[0.5ex] 
\hline 
Lower limit & 13.8 & 13.8 & 13.8 & 13& 12 & 8.1 & 6 & 6 \\
Upper limit & 21.2 & 20.6 & 20.3 &19.8 &19.6  & 18.5 & 17.7 & 18 \\[0.5ex] 
   \hline
   \hline
\end{tabular}
\label{tab:missing_phot}
\end{table*}

The photometric calibration was performed only for the $g,r,i,z,y$ and $J,H,Ks$ images. It was not attempted for the INT images obtained in any other filter, the ESO2.2m WFI images, the PTF images (because the camera has a significantly coarser pixel scale and the images reach a depth shallower than Pan-STARRS), and the CPAPIR $I$-band images.

The photometric zeropoint of all individual images was computed by direct comparison of the instrumental {\sc SExtractor} \verb|MAG_AUTO| magnitudes with an external catalogue: 
\begin{itemize}
\item $J,H,Ks$ images were tied to the 2MASS catalogue,
\item $g,r,i,z,y$ images were tied to the Pan-STARRS PS1 first release.
\end{itemize}
The procedure followed to derive the individual zeropoints is described in \citet{Olivares+19}. Briefly, the zeropoints are computed as the median of the difference between the instrumental magnitude and the measurements of the closest match within 1\arcsec\, in the reference catalogue after rejecting outliers using the modified Z-score criterion. We find typical 1$\sigma$ dispersions of the order of 0.03--0.08~mag depending on the filter.

We median-combined all the images obtained with the same camera and in the same filter to build a deep stack and extracted the corresponding photometry. This allowed us to significantly improve the sensitivity in all filters and recover or improve the photometry of faint sources obtained in the individual images. 

As in \citet{Bouy+13}, we complemented the photometry extracted from the images with that of external catalogues: \textit{Gaia} DR2 ($G$, $G_{\rm BP}$, $G_{\rm RP}$), Pan-STARRS ($grizy$), 2MASS ($JHKs$) and ALLWISE (all 4 bands) to improve the spatial and wavelength coverage of the final dataset (see Appx.~\ref{app:query_GDR2}--\ref{app:query_WISE} for the queries used). The corresponding photometric measurements were either added to our catalogue when no measurement was available in our data, or the weighted average of all measurements (our and external catalogues) was computed after rejecting outliers using the modified Z-score criterion. 

The differences between the transmission of the various instruments contribute to the photometric dispersion (e.g. see Fig.~\ref{fig:iband}). Moreover recent variability studies of young clusters found typical amplitudes of 0.03~mag \citep[e.g.][]{Rebull+16,Rebull+18}. As explained in \citet{Olivares+19} we add quadratically 0.05~mag to all the photometric measurement uncertainties in our catalogue to take into account these sources of uncertainties in our membership analysis. The final mean uncertainties in photometry depend on the photometric band, and are of the order of 0.07~mag.

Given the low extinction in that area of the sky, the maximum of the magnitude distributions (Fig.~\ref{fig:completeness}) gives an estimate of the completeness limit of the survey. It is nevertheless important to remember that the spatial coverage of the various instruments is not homogeneous and the depth of the survey varies spatially. The limit of sensitivity of the external catalogues merged with our data (2MASS and Pan-STARRS) are sometimes visible as secondary maxima. In Table~\ref{tab:missing_phot} we give, for each photometric band, the number and percentage of measurements as well as the completeness limits we use in this study.
This final catalogue (hereafter the DANCe catalogue) includes 2\,358\,937 sources.

\section{Membership analysis}
\label{sec:membership}

To select candidate members we used the methodology originally developed by \cite{Sarro+14} and updated by \cite{Olivares+19}. Briefly, this algorithm separates all the sources within two populations, namely the cluster and the field. The field model is computed once at the beginning and fixed thereafter, based on the assumption that the few hundreds of cluster members do no affect significantly the field population model. The cluster model is built iteratively and, at each iteration, both models are used to reclassify the sources until convergence. This algorithm takes into account the covariance matrix of the astrometric parameters when available. To model the cluster and field populations, the algorithm only uses the complete sources, i.e. the sources with measurements available for all the observables. The  coverage and sensitivity of the different photometric bands is therefore a key issue in our analysis. The final model allows us to compute a membership probability to incomplete sources after marginalisation over the missing values. To start the analysis and build the first model we need a catalogue of sources for the region of interest and an initial list of members. The latter can be slightly contaminated and incomplete, and serves only to define the cluster locus in the multi-dimensional space in the first iteration. 

Because the field and cluster models are built from sources with complete photometric and astrometric measurements, a simultaneous analysis of the GDR2 and DANCe catalogues would not be optimum. Many faint DANCe sources would have missing parallaxes and \textit{Gaia} photometry while bright GDR2 sources, saturated in our images, would have missing DANCe photometry, making it hard (if at all possible) to define a proper representation space in which a sufficient number of sources have complete measurements. As described in \citet{Olivares+19} we therefore decided to analyse both catalogues independently.

\subsection{Initial members}
Recently, two studies have published members of IC~4665 using the \textit{Gaia} DR2 data. The work of \cite{GaiaColBabusiaux+18} published a list of 174 members and the work of \cite{Cantat-Gaudin+18} published a list of 175 members. Both studies have a magnitude limit of $G=18$, and most of the sources in common. We combined their results and obtained a list of 203 members which we used as initial list for our membership analysis of the GDR2 catalogue. 

To start the membership algorithm for the DANCe catalogue, we used the members we obtained with the GDR2 catalogue which have a counterpart in DANCe. In this case, the initial list does not cover the full magnitude range of the catalogue (DANCe goes fainter than the initial list of GDR2 members). However, our algorithm extends the initial principal curve (see the cluster model for the photometry in Sect~\ref{subsec:cluster-model}) by progressively and iteratively extrapolating the photometric sequence to fainter regions. These fainter regions of extrapolation are small and the new candidate members found with them are added and used in subsequent iterations to better define (or correct if necessary) the extrapolation. The extrapolation of the photometric curve is guided by the astrometry which does not change with magnitude. This extrapolation of the principal curve is further explained in \citet{Sarro+14} where we presented for the first time the membership algorithm.

\subsection{Representation space}
\label{subsec:RS}
The representation space is the set of astrometric and photometric variables we use for the membership analysis. We always use proper motions, as well as parallaxes in the case of the GDR2 catalogue. The photometric variables are chosen according to their importance calculated from a random forest algorithm, as in \citet{Olivares+19}. The largest the amount of features, the more information there is to classify the sources between members and non-members. At the same time, the bands with a large number of missing observations are avoided. 

With the representation space established, the whole data set is split between complete and incomplete sources. A source is said to be complete when it has a measurement for all the variables of the chosen representation space. In consequence, different representation spaces lead to different complete/incomplete ratios.

For the analysis with the GDR2 catalogue, the representation space we used is \texttt{pmra}, \texttt{pmdec}, \texttt{parallax}, $G_{\rm RP}$, $G_{\rm BP}-G$, $G-G_{\rm RP}$. With this representation space 1\,184\,922 sources (97\%) have complete data.

For the analysis with the DANCe catalogue, the representation space we used is \texttt{pmra}, \texttt{pmdec}, $J$, $i-z$, $i-y$, $i-J$. With this representation space, 1\,627\,593 sources have observations in all the photometric bands, which represents a 69\% of the catalogue. We decided not to include the \textit{g}, \textit{r}, \textit{H}, and \textit{$Ks$} bands in the representation space because of the large number of sources with missing photometry (see Table~\ref{tab:missing_phot}).

\subsection{Field model}

The field population is modelled with a Gaussian Mixture Model (hereafter GMM) in the whole representation space. The field model is computed once at the beginning and fixed thereafter. We explored GMMs with different number of components (60, 80, 100, 120, 140, 160, and 180). We choose the optimum model as the simplest one which minimises the Bayesian information criterion (BIC). This results in 100 components for the analysis with the GDR2 and the DANCe catalogue.

\subsection{Cluster model}
\label{subsec:cluster-model}

The cluster model is a product of two independent models: a GMM for the astrometry and a principal curve in photometry. The astrometric model is a GMM and we choose the number of components that optimises the BIC between 1 and 4.

The cluster model is computed iteratively, starting from the initial list of members. At each iteration, we compute independently a model for the astrometry and a model for the photometry. Then, we assign Bayesian probabilities of membership to all the complete sources. These probabilities together with a probability threshold, $p_{in}$, are used to reclassify the complete sources between members and non-members. The $p_{in}$ is a free parameter of the model which defines the degrees of completeness and contamination that we desire for the training set (and as a consequence, for the final list of members). We refer the interested reader to \citet{Sarro+14} and \citet{Olivares+19} for a more detailed description of this parameter. Then, the cluster model is recomputed based on the new members list and we repeat this process until convergence. 

When the model has converged, we generate a synthetic dataset from the model learnt with observed data. Therefore it has similar properties as the latter (e.g. missing values, frequency of members). As a consequence, the results derived from them are restricted to the used RS and learnt model.

We use this synthetic dataset to analyse the goodness of our classification as well as to compute the optimum probability threshold, $p_{opt}$, which is used to do a final classification. The optimum threshold will of course depend on the scientific goal behind the membership analysis. In our case, to study the mass function, we are interested in reaching a compromise between the contamination and the true positive rate. For that, we choose as $p_{opt}$ the one which minimises the distance to the perfect classifier (DST). This distance is defined in terms of the contamination rate (CR) and the true positive rate (TPR) which in turn depend on the the confusion matrix: true positives (TP), false positives (FP), false negatives (FN), and true negatives (TN). This indices are defined as follows:
\begin{align*}
CR  &= \frac{FP}{FP+TP} \\
TPR &= \frac{FP}{FP+TP} \\ 
DST &= \sqrt{(\textup{CR-0})^2 + (\textup{TPR}-1)^2}
\end{align*}

As we have mentioned, the estimations which can be obtained with this synthetic data set are restricted to the same conditions as the observations and to the assumption that the model correctly represents the observed data. Thus, the measured CR and TPR can be underestimated and overestimated respectively, with respect to those obtained with better quality data and realistic models. As we shall see in Sect.~\ref{subsec:final_memb_list}, this is observed in the CR of DANCe. The value estimated using synthetic data appears to be underestimated with respect to the one obtained using the GDR2 members as reference. For more details we refer to \citet{Olivares+19}.

We run the full model considering several $p_{in}$ thresholds (0.5, 0.6, 0.7, 0.8, and 0.9) and, for each, we compute the optimum threshold $p_{opt}$, using synthetic data. In Table~\ref{tab:pin} we show the $p_{in}$, $p_{opt}$, the number of complete and incomplete members, for each of the two independent analysis (GDR2 and DANCe). 

\begin{table}
\caption{Internal probability threshold ($p_{in}$), optimum probability threshold ($p_{opt}$) and number of complete and incomplete members (CM, IM) for each of the membership analyses, namely GDR2 and DANCe.  }
\centering
\begin{tabular}{|c||ccc|ccc|}
\hline
\hline

         & \multicolumn{3}{c|}{GDR2} & \multicolumn{3}{c|}{DANCe}  \\ \hline \hline
$p_{in}$ & $p_{opt}$ & CM    & IM  & $p_{opt}$ & CM    & IM  \\ \hline  
0.5      & 0.86      & 539   & 0   & 0.86      & 708   & 8   \\    
0.6      & 0.78      & 567   & 0   & 0.87      & 665   & 4   \\    
0.7      & 0.77      & 434   & 0   & 0.87      & 643   & 4   \\    
0.8      & 0.76      & 405   & 0   & 0.83      & 639   & 5   \\    
0.9      & 0.68      & 383   & 0   & 0.80      & 578   & 4   \\ \hline \hline   
\end{tabular}
\label{tab:pin}
\end{table}

\subsection{Classification of incomplete sources}

We used the field and cluster models described in the previous subsections to compute membership probabilities for the incomplete sources, i.e. the sources which lack one or more magnitudes of the representation space. Then, we used the optimum threshold to classify all the sources between members and non-members.

For the GDR2 catalogue, there are very few incomplete sources and none of them is classified as member. For the DANCe catalogue, the number of incomplete sources classified as members is 4--8 depending on the $p_{in}$. In general, they lack $z$ and/or $y$ photometry and the brightest ones are also classified as members by the analysis with GDR2.

\subsection{Final members lists}
\label{subsec:final_memb_list}

\begin{table*}
\caption{GDR2 catalogue of IC4665 (only the first 10 rows are displayed as example). Columns indicate: (1) \textit{Gaia} DR2 source ID; (2--3) right ascension and declination; (4) \textit{Gaia} DR2 $G$-band magnitude; (5--9) posterior probabilities obtained with $p_{in}$ from 0.5 to 0.9. }
\centering
\begin{tabular}{ccccccccc}
\hline
\hline
source ID & RA      & Dec     & $G$   & $p_{in}=0.5$ & $p_{in}=0.6$ & $p_{in}=0.7$ & $p_{in}=0.8$ & $p_{in}=0.9$ \\
          & [\degr] & [\degr] & [mag] &              &              &              &              &              \\
\hline
  4376249758236481664 & 265.96 & 2.82 & 20.35 & 2.5E-29 & 7.2E-31 & 8.02E-30 & 7.2E-37 & 8.8E-45\\
  4376260409754607872 & 265.81 & 2.80 & 19.27 & 9.2E-45 & 6.4E-39 & 1.75E-42 & 3.9E-58 & 1.8E-58\\
  4376260482773352576 & 265.80 & 2.80 & 18.10 & 1.4E-51 & 3.0E-54 & 3.67E-60 & 7.5E-67 & 6.4E-80\\
  4376260482769778176 & 265.80 & 2.80 & 20.28 & 3.1E-18 & 6.9E-20 & 2.89E-21 & 1.0E-22 & 2.5E-24\\
  4376260478474097664 & 265.80 & 2.80 & 18.62 & 3.4E-61 & 8.3E-56 & 1.34E-79 & 9.0E-97 & 1.2E-107\\
  4376260650272889856 & 265.78 & 2.81 & 19.29 & 6.5E-5  & 4.0E-5  & 8.33E-6  & 4.9E-6  & 1.4E-6 \\
  4376260684632540032 & 265.78 & 2.81 & 19.36 & 1.6E-32 & 6.1E-34 & 1.31E-35 & 3.0E-46 & 1.6E-48\\
  4376260684632546816 & 265.78 & 2.81 & 18.12 & 1.7E-62 & 1.4E-43 & 2.3E-56  & 5.9E-61 & 1.7E-71\\
  4376260684632548736 & 265.79 & 2.81 & 19.03 & 3.7E-39 & 3.4E-40 & 1.82E-45 & 2.6E-55 & 4.6E-58\\
  4376260684633273344 & 265.79 & 2.82 & 19.82 & 2.8E-11 & 7.1E-12 & 3.89E-12 & 2.1E-15 & 5.6E-18\\
\hline
\hline
\end{tabular}
\label{tab:GDR2_cat}
\end{table*}
\begin{table*}
\caption{DANCe catalogue of IC4665 (only a subset of the most relevant columns, and the first 10 rows are displayed as example). Columns indicate: (1--2) right ascension and declination; (3--4) proper motions; (5--12) photometry; (13 and 15) posterior probabilities obtained with $p_{in}$ of 0.5 and 0.9. }
\centering
\renewcommand\tabcolsep{4pt}
\begin{tabular}{ccccccccccccccc}
\hline
\hline
 RA      & Dec     &pmRA            & pmDec           & $g$   & $r$   & $i$   & $z$  & $y$     & $J$     & $H$     & $Ks$     & $p_{in}=0.5$ & ... & $p_{in}=0.9$ \\
 
 [\degr] & [\degr] &[mas yr$^{-1}$] & [mas yr$^{-1}$] & [mag] & [mag] & [mag] & [mag] & [mag] & [mag] & [mag] & [mag] &              &      &  \\
\hline         
266.37 & 5.82 & - 4.51 & - 7.98 & 19.7 & 19.8 & 19.0 & 19.0 & 19.0 & 17.8 & 17.0 & 17.0 & 6.3E-54 & ...  & 7.1E-65\\
266.67 & 5.73 & -10.19 & -23.17 & 19.8 & 19.2 & 19.0 & 18.9 & 18.9 & 17.9 & 17.5 & 17.3 & 3.3E-86 & ...  & 3.2E-119\\
266.30 & 5.67 &   5.99 & - 8.06 &      & 21.0 & 19.7 & 19.0 & 18.7 & 17.3 & 16.7 & 16.4 & 1.7E-18 & ...  & 9.8E-20\\
266.39 & 6.90 &   0.81 & - 4.07 & 20.4 & 19.8 & 19.3 & 19.1 & 19.3 &      &      &      & 7.1E-6  & ...  & 2.4E-8\\
266.66 & 6.81 & - 2.51 & - 5.41 & 20.4 & 17.8 &      & 16.1 &      &      &      &      & 1.2E-4  & ...  & 4.9E-5\\
266.86 & 6.74 & - 6.60 & - 8.79 &      &      &      &      &      &      &      &      & 3.2E-6  & ...  & 1.3E-7\\
266.86 & 6.79 &   1.62 & - 8.71 & 20.4 & 19.9 & 19.6 & 19.4 & 19.2 & 18.2 & 17.5 &      & 6.2E-62 & ...  & 1.5E-69\\
266.89 & 6.92 &   2.66 & - 9.60 & 18.5 & 18.0 & 17.8 & 17.7 & 17.7 & 16.6 & 16.2 & 16.2 & 5.7E-51 & ...  & 2.2E-67\\
266.21 & 6.63 & - 5.09 & - 5.88 & 19.6 & 19.0 & 18.8 & 18.7 & 18.7 & 17.6 & 17.2 & 17.1 & 1.3E-55 & ...  & 6.0E-85\\
266.89 & 6.92 & - 2.22 & -11.55 & 18.9 & 18.1 & 17.8 & 17.6 & 17.4 & 16.3 & 15.9 & 15.7 & 5.7E-29 & ...  & 8.4E-38\\
\hline
\hline\end{tabular}
\label{tab:DANCe_cat}
\end{table*}

Tables~\ref{tab:GDR2_cat} and \ref{tab:DANCe_cat} (available at CDS) give the GDR2 and DANCe catalogues used in this work, respectively. For the GDR2 catalogue, we provide the \textit{Gaia} DR2 source ID and the sky positions, and for the DANCe catalogue, we compile all the astrometry and photometry presented in Sect.~\ref{sec:data}. In addition, for each catalogue, we provide the posterior membership probabilities obtained with the different $p_{in}$ discussed in this section. Here, we describe our strategy to choose the members list most convenient to our necessities. However, we encourage the interested reader to chose the members list most convenient to his/her goal.

The membership probabilities obtained with different $p_{in}$ values have to be compared with care. The relation between different membership probabilities (obtained with different $p_{in}$) is not linear, and lower $p_{in}$ tend to provide higher membership probabilities. In general, the models computed with lower $p_{in}$ permit a larger inclusion of sources initially classified as field into the cluster class during the training of the model. This results in lists of members which can include a significant amount of contaminants. On the contrary, models computed with higher $p_{in}$ are more restrictive in including additional sources into the cluster model and thus, tend to have lower contamination but at the same time can become incomplete. 

The membership probabilities we compute are not absolute but tightly related to the model used which at the same time depends on the representation space and on the $p_{in}$ (desired degree of completeness and contamination). In consequence, the comparison between GDR2 and DANCe membership probabilities is not straightforward, specially due to the different representation spaces of each catalogue (DANCe does not have parallaxes). 

To study cluster properties such as the mass function, we want a unique list of members, the cleanest and most complete possible. This means first, choosing a $p_{in}$ for each study (GDR2 and DANCe), and then combining both catalogues. We begin with the analysis of GDR2, which is expected to be more robust since it includes a very discriminating variable: the parallax. We used a Kolmogorov-Smirnov (KS) test in different variables of the representation space (proper motions, parallax, and photometry) to see if the distributions of these variables obtained with different $p_{in}$ were compatible among them. The goal is to see if we find signs of a strong contamination or a strong incompleteness in one or several of the lists with respect to the others. We started by taking the sources classified as members (i.e. those with $p>p_{opt}$) obtained with the model trained with $p_{in}=0.9$ as a reference. This is the most conservative and the least contaminated, but also probably the most incomplete list of members. Then, we compared this list of members with all the rest ($p_{in}=0.5$, 0.6, 0.7, and 0.8, one at each time). The KS test showed that for the lists of $p_{in}=0.5$ and 0.6, there is evidence that the distribution of their proper motions and parallax do not come from the same distribution as the one obtained with the list of $p_{in}=0.9$ with a p-value lower than the significance level $0.01$. On the contrary, for the distribution of the astrometric variables coming from the lists of $p_{in}=0.7$ and 0.8, the KS test shows no evidence to reject they come from the same distribution as the one obtained with $p_{in}=0.9$ with p-values of 0.4--0.5. Then, we investigated which is the reason of the incompatibility of the lists of $p_{in}=0.5$ and 0.6 with respect to the rest. We found that the parallax and proper motions distributions obtained with the lists of $p_{in}=0.5$ and 0.6 have significantly more extended wings than the distributions obtained with the lists of higher  $p_{in}$. We interpreted this as contamination and therefore, we discarded these two lists. The remaining lists are compatible according to the KS test so we chose the list of $p_{in}=0.7$ which has the largest number of members. 

To select the optimum $p_{in}$ for the DANCe analysis we also applied a KS test to find which distributions were compatible with the one obtained with $p_{in}=0.9$. In this case, there was no evidence to reject that the distributions of all the variables analysed for the lists from all $p_{in}$ come from the same distribution as the ones of $p_{in}=0.9$ since all the p-values were $>0.3$. Then, to check the consistency between the GDR2 and DANCe lists, we took the members of GDR2 $p_{in}=0.7$ as a reference and compared them with the members recovered in the different DANCe lists, in the region where both studies are complete ($14.5\lesssim G\lesssim19$). We found that all the DANCe lists recovered roughly the same number of GDR2 members (250--266 from 285, $\sim90\%$). On the contrary, the number of members in DANCe which were not in GDR2 decreased with increasing $p_{in}$. These sources have parallaxes which are in general incompatible with the GDR2 members (further than $3\sigma$), and thus we believe that most of them are contaminants (they represent a 30--35\% of the DANCe members). In short, for the DANCe analysis we do not find any strong argument to discard any list. Therefore, we decided to keep the one with largest amount of members, i.e. the one with $p_{in}=0.5$ keeping in mind that it includes a contamination of the order of 30--35\%, estimated from the comparison with the GDR2 members.

In the rest of this work, we will be using the GDR2 list with a $p_{in}=0.7$ and the DANCe list of $p_{in}=0.5$. Those two lists add to a final list of 819 members which is analysed in more detail in the next section.

\section{Members comparison}
\label{sec:analysis-members}
In this section we analyse and discuss our final members list. We compare the results obtained with the GDR2 and DANCe catalogues, and we also compare our final list with the members already reported in the literature.

\subsection{Comparison GDR2 vs. DANCe members}
\begin{figure}
\begin{center}
\includegraphics[width = \columnwidth]{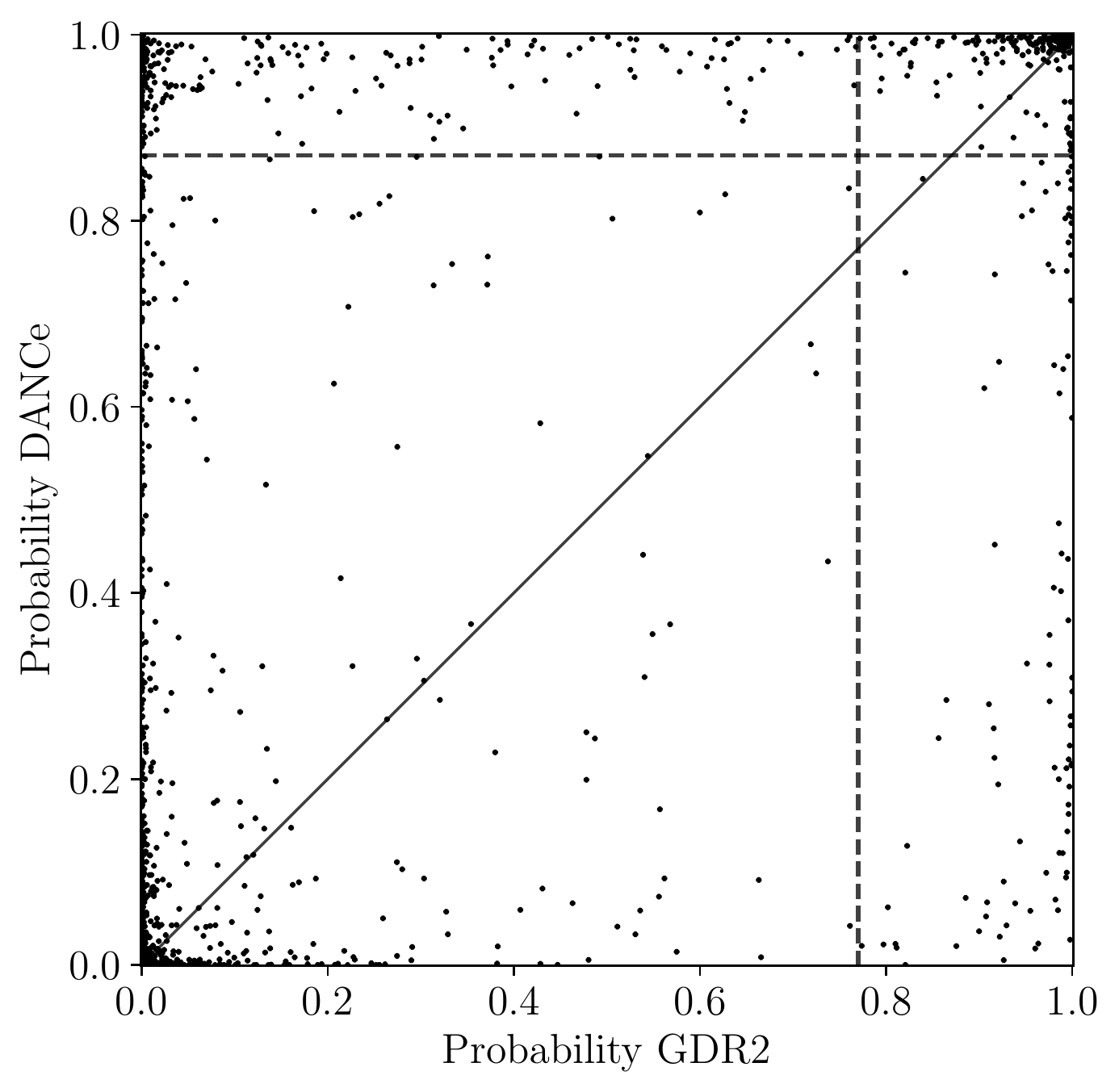}
\caption{Comparison between the membership probabilities recovered by the GDR2 classifier and the DANCe one, for objects in both catalogues. The diagonal line represents the one-to-one relation, and the horizontal and vertical dashed lines show the optimum probability thresholds.}
\label{fig:probabilities}
\end{center}
\end{figure}

\begin{figure}
\begin{center}
\includegraphics[width = \columnwidth]{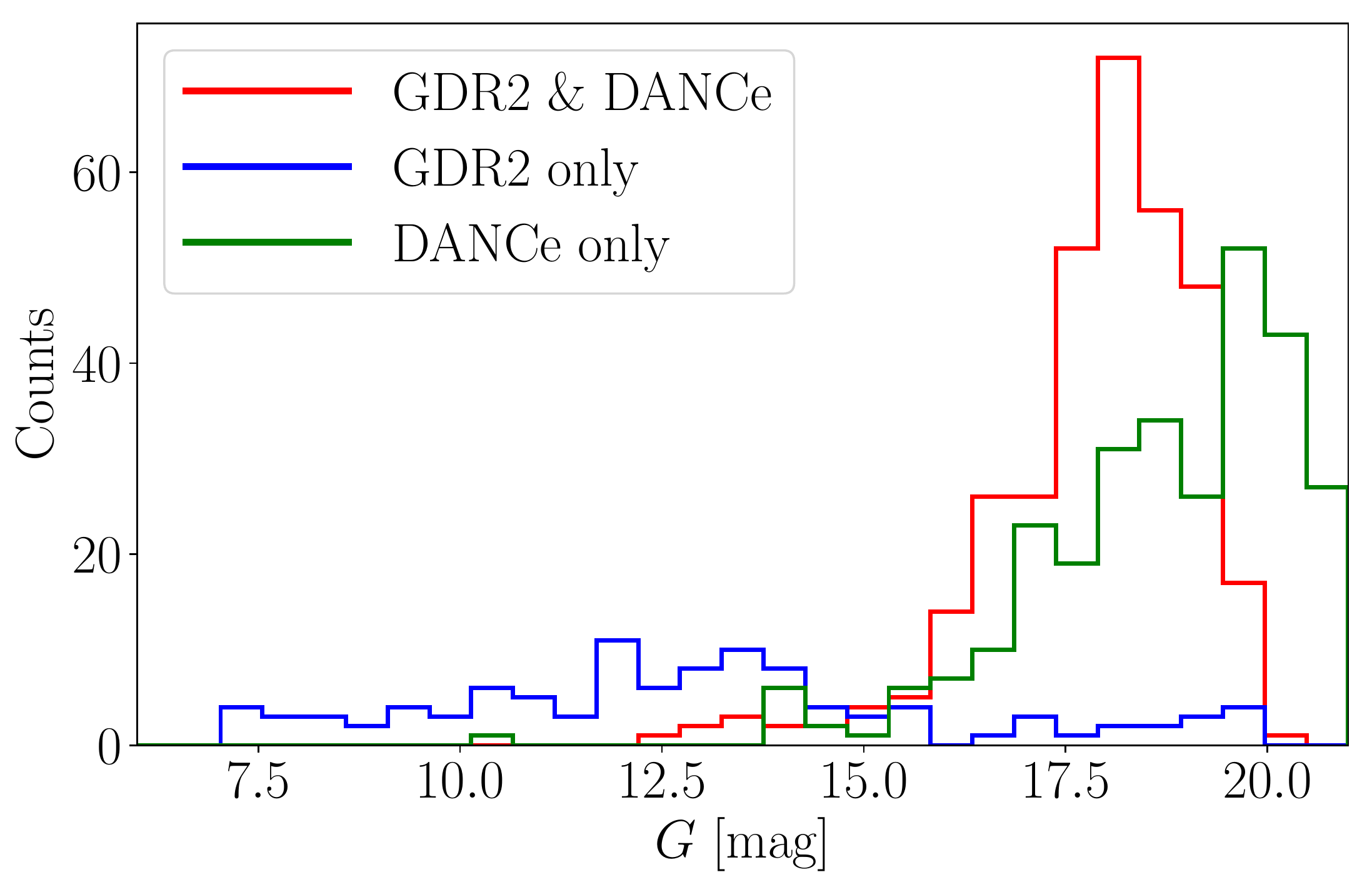}
\includegraphics[width = \columnwidth]{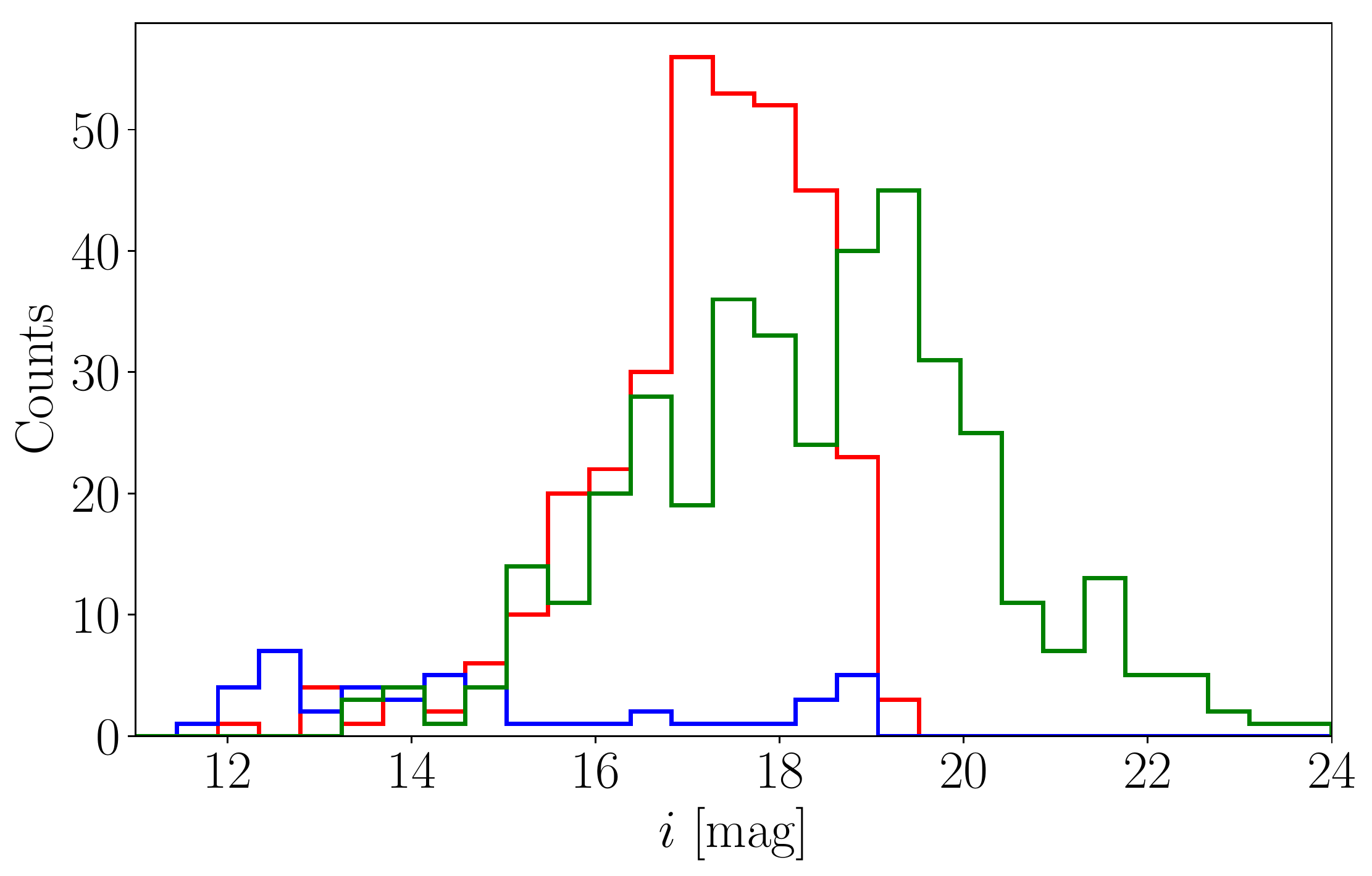}
\caption{$G$ and $i$ (top and bottom) magnitude distributions of the sources classified as members in the GDR2 and DANCe studies (red), classified as members by GDR2 but not by DANCe (blue) and classified as members by DANCe but not by GDR2 (green). }
\label{fig:mag-DANCeGaia}
\end{center}
\end{figure}

\begin{figure*}
\begin{center}
\includegraphics[width =1\textwidth]{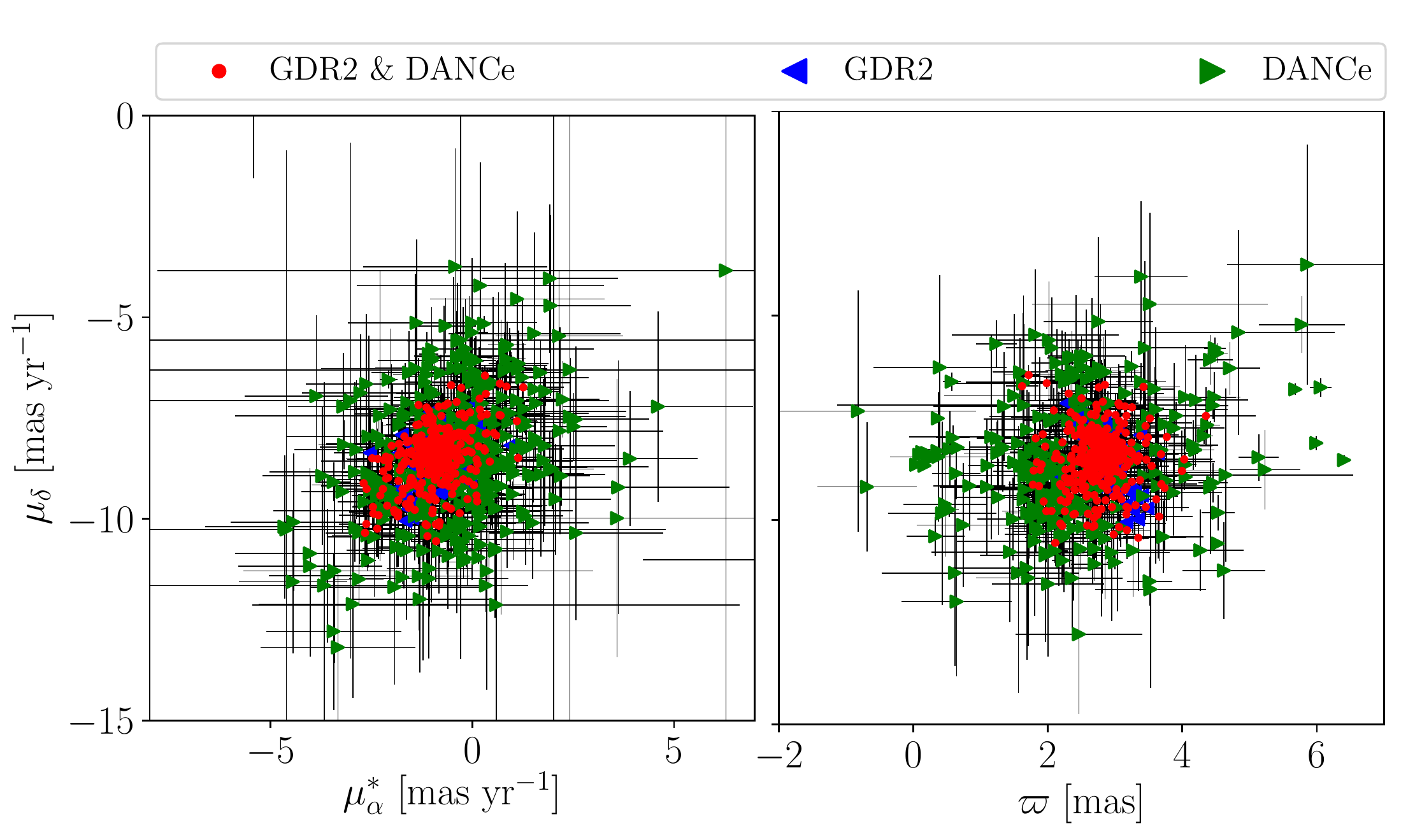}
\caption{Vector point diagram (left) and parallax--proper motion diagram (right) of the IC~4665 open cluster. The members are shape and colour coded according their origin: GDR2 and DANCe analysis (red circles), only the GDR2 analysis (blue left-pointing triangle) and only the DANCe analysis (green right-pointing triangle). }
\label{fig:astrometry}
\end{center}
\end{figure*}

Here we compare the two membership analyses obtained with the GDR2 and the DANCe catalogues. We cross-matched the two catalogues (which contain both members and field stars) and found 1\,211\,272 sources in common. In Figure~\ref{fig:probabilities} we compare the membership probabilities obtained with the two catalogues. The diagonal line represents the one-to-one relation, and the vertical and horizontal dashed lines represent the optimum thresholds. We see that most of the sources are clustered in the bottom left (field) and top right (cluster) regions of the diagram. Nonetheless, there are sources which are classified as members by one of the two studies and not by the other. 

To understand the differences between the two classifiers we represented the number of members as a function of the magnitude (Fig.~\ref{fig:mag-DANCeGaia}). We distinguish between the members obtained with both classifiers (red), the members only from the GDR2 analysis (blue), and the members only from the DANCe analysis (green). Here we discuss the 4 possible cases regarding the results of the two membership analyses. 

\subsubsection{Members in GDR2 and DANCe}
There are 331 sources which appear as members in both analyses (red in Fig~\ref{fig:mag-DANCeGaia}). In the magnitude range where both catalogues are complete, the majority of members are classified as so by the two analyses.
 
\subsubsection{Members in GDR2 only}
There are 103 sources which appear as members in the GDR2 analysis but not in DANCe (blue in Fig~\ref{fig:mag-DANCeGaia}). The majority of these are the brightest sources which are saturated in the DANCe catalogue. Most of them have $J$, $H$ and $Ks$ photometry but lack $i$, $z$ and $y$ which are essential bands for the representation space we use in DANCe. 

In the magnitude range where both analyses are complete, we see that the members obtained only with GDR2 have a low and flat distribution. This can be interpreted as the GDR2 members list having a very low contamination, which does not depend on the magnitude. The reason is that the parallax is the most discriminating variable to classify these members. Although the uncertainties on the parallax depend on the magnitude, they are at the level required to distinguish the cluster from the field along all the magnitude range. At magnitudes $>18$ we see a slight increase but it is not significant.

\subsubsection{Members in DANCe only} 
There are 385 sources which appear as members in DANCe but not in GDR2 (green in Fig~\ref{fig:mag-DANCeGaia}). From these, 120 objects (31\%) do not have the five-parameter solution in Gaia and 186 (48\%) have parallax uncertainties $>10\%$. Aside, we discussed in the previous section that we find a $\sim30\%$ of contamination in the region where both studies are complete. This is significantly larger than what we found in other clusters (i.e. the Pleiades and Ruprecht 147, \citealp[see][respectively]{Sarro+14, Olivares+19}) but this is expected given the lower galactic latitude and significantly smaller proper motions of IC~4665. 

In addition, we see that the amount of members only recovered by DANCe increases as a function of magnitude in the region where both analyses are complete. We interpret this as a dependence of the contamination on the magnitude. The DANCe analysis does not use the parallax and thus it is expected that the photometry plays a major role, specially in this cluster with small proper motions.

\subsubsection{Non-members in GDR2 nor in DANCe} 
All the remaining sources are classified as field stars by both studies. Most of them have extremely low membership probabilities which clearly identifies them as field population. There are several sources which have rather high probabilities but fall below the threshold. This means we can not definitely discard them as members and could be considered as candidate members depending on the scientific case. The sources which are spread along the rest of the diagram may suffer from the problems already discussed or simply, the observables in the two catalogues are too different. To clarify the membership of the uncertain cases we would need either a longer temporal base-line to improve the proper motions or spectroscopy to study their properties (i.e. low gravity due to youth).

\vspace{5mm}

In short, we see that in general the two independent analysis agree rather well, specially in the magnitude range were both are expected to perform well. The members obtained with both catalogues occupy the same space in the vector point diagram (see Fig.~\ref{fig:astrometry} left). The members coming from the DANCe catalogue typically have a larger dispersion and larger uncertainties, expected by the different precision of both catalogues. In the space of parallaxes (Fig.~\ref{fig:astrometry} right) we see that the members from the GDR2 analysis are very well concentrated around the median value (2.84~mas with a standard deviation of 0.36~mas). Some members from the DANCe analysis, which do not use this parameter for the classification, also have parallaxes compatible with the cluster distribution. Others have parallaxes incompatible with the cluster (at 3$\sigma$ level) and they are either problematic measurements (because they are very faint) or contaminants. Future releases of the \textit{Gaia} catalogue will help to clarify these cases. 

When we introduced the GDR2 catalogue in Sect.~\ref{sec:gaiadr2}, we mentioned that we have not filtered the data in any manner in order to be the most complete possible. Here, we discuss the RUWE goodness of fit indicator of the members found in this study. Our sample contains sources with a RUWE in the range  0.8--16.3, and only 9\% of them have a RUWE larger than the recommended threshold (1.40). However, we insist that all the sources with a RUWE larger than the recommended one do not always have a wrong solution and future releases of \textit{Gaia} or complementary observations will tell. 

\subsection{Comparison with other studies}

\begin{figure}
\begin{center}
\includegraphics[width = \columnwidth]{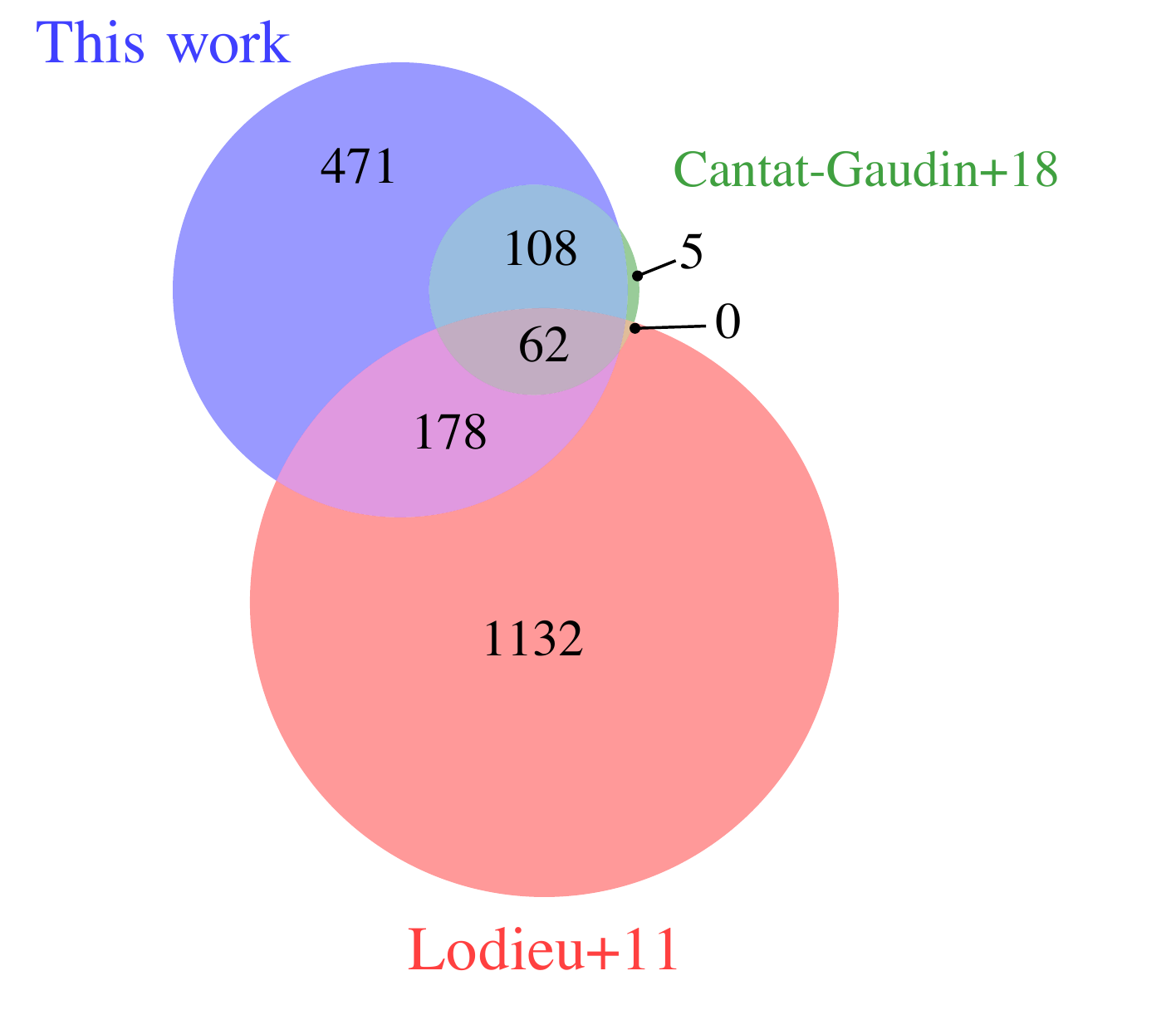}
\caption{Venn diagram comparing the members in this work to previous studies in the literature \citep[i.e.][]{Lodieu+11, Cantat-Gaudin+18}.}
\label{fig:venn_diagram}
\end{center}
\end{figure}

In this section we compared our list of 819 members with other studies in the literature and found that 409 (50\%) are new members. We have cross-matched each of the members lists reported in the literature with ours using a maximum separation of 1\arcsec. 

In Figure~\ref{fig:venn_diagram} we compare the members we find with two of the most representative membership studies of IC~4665: the one from \citet{Lodieu+11} based mainly on photometry, and the one from \citet{Cantat-Gaudin+18} using the \textit{Gaia} DR2 astrometry. As a general trend, purely photometric studies tend to have more contamination than spectroscopic ones or the ones based on \textit{Gaia} DR2 astrometry.

\subsubsection{\citet{deWit+06}}
These authors photometrically selected 691 low-mass stellar and 94 brown dwarf candidate members over an area of 3.82 square degrees centred on the cluster. In addition, they applied a filter for bright stars based on the proper motions from Tycho-2 and UCAC2 public catalogues. 

We detected some astrometric offsets between their positions and ours and consequently extended the cross-match search radius to 2\arcsec. We confirmed 195 of their members and rejected the rest of their candidates which have very low membership probabilities in our study. Therefore, we estimate a contamination up to 75\% in their study compatible with their own estimate. We believe that one of the reasons of their large contamination is a problematic photometric calibration (their \textit{i} and \textit{z} bands photometry display an offset of $\sim1$ mag compared to Pan-STARRS).

\subsubsection{\citet{Manzi+08}}
These authors did not attempt to do a comprehensive census of the cluster. Instead, they photometrically selected candidates from the literature and then spectroscopically confirmed 37 of them. Their aim was to determine the age of IC~4665 using the lithium depletion boundary method. We confirmed 29 of their members (78\%) and discarded the remaining 8 (two of them were classified as not fully secure members by the authors). These 8 members are discarded because their \textit{Gaia} DR2 parallaxes and/or proper motions are far from the cluster distribution, although their photometry falls on the cluster sequence. Therefore, these sources are either interlopers or have a problematic astrometric solution in \textit{Gaia} DR2.

\subsubsection{\citet{Jeffries+09}}
These authors aimed to study the pre-main-sequence lithium depletion for low-mass stars in IC~4665. For this purpose, they selected 40 members according to several spectroscopic criteria. We confirmed 30 of their members (75\%) and rejected the remaining 10. This study has 12 members in common with \citet{Manzi+08} and from those, only one source is rejected by our study. Again, the 10 members are rejected by our analysis because their \textit{Gaia} DR2 astrometry is incompatible with that of the cluster, but the same reasoning of \citet{Manzi+08} applies.

\subsubsection{\citet{Cargile+10}}
These authors used a photometrically selected sample of members in the central region of the cluster (1 square degree) to study the age and distance of IC~4665. Their sample contained 382 candidates members, 49 of which are confirmed by our study. From this, we estimated their contamination to be 87\%.

\subsubsection{\citet{Lodieu+11}}
These authors used photometry from UKIDSS and CFHT to identify members in IC~4665. They presented a sample of 1\,372 members in the magnitude range $15<i<20.4$ which they used to study the luminosity and mass functions.  

Only 240 of their candidates (17\%) were classified as members in our work (see Fig.~\ref{fig:venn_diagram}). The majority of the rejected candidates have extremely low membership probabilities in our analysis (both in GDR2 and DANCe). We believe the reason of their large contamination ($\sim$80\%) is the same photometric offset as for \citet{deWit+06} since they used the same data.

These works constituted the most exhaustive study, specially regarding the low-mass regime, previous to the results we present here. Given the high levels of contamination found by the present analysis, we hereafter do not attempt any comparison of their luminosity and mass function.

\subsubsection{\citet{Bravi+18}}
These authors used the Gaia ESO Survey to study the IC~4665 open cluster. They carried out spectrosopic observations of 567 sources in the region of the cluster. They used spectroscopic criteria to exclude obvious contaminants and then, they computed membership probabilities using the radial velocity distribution of the cluster and the field. They ended up with 29 sources with membership probability $>0.5$, 24 of which greater then $>0.8$. From these 29 candidates, 20 were confirmed by our study (15 have probabilities $>0.8$ according to their study), and the remaining 9 were definitely rejected in our study. As for the previous spectroscopic surveys, the reason we discard these 9 members is the \textit{Gaia} DR2 astrometry, which is incompatible with that of the cluster.

\subsubsection{\citet{GaiaColBabusiaux+18}}
With the purpose of demonstrating the power of \textit{Gaia} DR2 in highlighting the fine structures of the Hertzsprung-Russell diagram, these authors selected members for a number of open clusters. Their ambitious goal requested to select only the sources with the highest precision in astrometry and photometry, and among other filters, they restricted to sources brighter than $G=18$. One of the clusters of their study is IC~4665, for which they provided a list of  174 members based only on the astrometric solution of \textit{Gaia} DR2. They claimed their list not to be complete but to contain potential members, i.e. to have an extremely low contamination rate. 

To do a fair comparison with this study, we only considered our members which are in the same magnitude and spatial range (brighter than $G=18$ and $2.4\degr$ radii around the centre of the cluster). This results in 267 members, 215 of which are classified as members by our analysis with the GDR2 catalogue and the rest come from the analysis with the DANCe catalogue only.

The study of \citet{GaiaColBabusiaux+18} and ours have 162 members in common which is a 93\% of their list. From the 12 objects classified as members by these authors and not by our study, there are 4 which have probabilities $>0.5$ but fall below the optimum threshold we adopted, and 8 which have lower probabilities. From these 8 members, only 1 was also classified as member by a similar study \citep{Cantat-Gaudin+18}. These small differences are part of the Poissonian noise of the membership analysis. In addition, we find 53 members not detected by these authors which are spread along all the parameter spaces (proper motions, parallax, and magnitude), following the cluster distribution. Some of these 53 members could have been discarded by the authors in their data filtering.

\subsubsection{\citet{Cantat-Gaudin+18}}
These authors provided a membership analysis for a large number of clusters making use of the recent \textit{Gaia} DR2 data. In order to avoid large uncertainties, they restricted to the sources brighter than $G=18$. They used an unsupervised membership algorithm to derive membership probabilities using only the astrometry of \textit{Gaia} DR2, and they found 175 members of IC~4665. 

To do a fair comparison with this study, we only considered our members which are brighter than $G=18$ and occupy the same spatial region of the sky ($\sim2\degr$ radius around the centre of the cluster). This results in 244 potential members, 205 of which come from the analysis of the GDR2 catalogue and the rest only from the analysis of the DANCe catalogue. 

The study of \citet{Cantat-Gaudin+18} and ours have 170 members in common which is a 97\% of their list. From the 5 objects classified as members by these authors and not by our study, there are 4 which have probabilities $>0.5$ but fall below the optimum threshold we adopted, and 1 which has a probability of 0.2. Again, these small differences are part of the uncertainties of the membership analysis. In addition, we find 35 members not detected by \citet{Cantat-Gaudin+18}, 17 of which are also classified as members by \citet{GaiaColBabusiaux+18}. These members are randomly distributed within the proper motions and parallax distributions. We found 3 very bright members of $G$ magnitudes 7.5, 9.5 and 10.5 and the rest are fainter than $G=14.5$. The DANCe members not classified by GDR2 in this magnitude range (39 sources with $14.5<G<18$) are likely to be contaminants, as discussed in Sect.~\ref{sec:membership}.

\section{Empirical and theoretical isochrones}
\label{sec:isochrones}

In this section we provide the empirical isochrones obtained with our members of IC~4665. Then we compare them with theoretical evolutionary models in several apparent colour magnitude diagrams. Finally, we convert the apparent colour magnitude diagrams to absolute colour magnitude diagrams and discuss them.

\subsection{Empirical isochrones}
\label{subsec:empirical_isoc}

\begin{figure*}[ht!]
\begin{center}
\includegraphics[width = 1\textwidth]{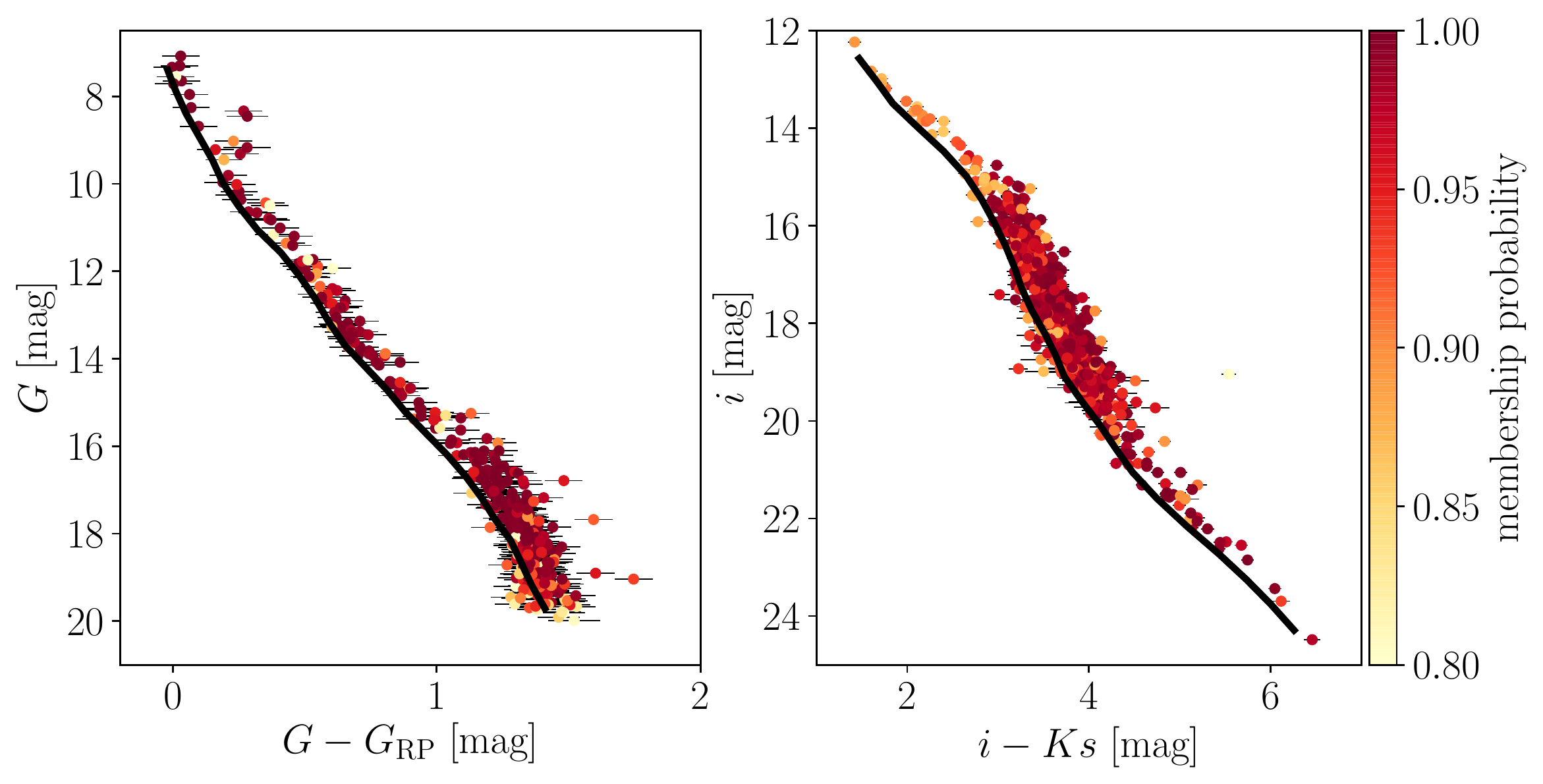}
\caption{($G$, $G-G_{\rm RP}$) and ($i$, $i-Ks$) colour magnitude diagrams (left and right, respectively) of the IC~4665 open cluster. The members are colour coded according to their membership probability and the empirical isochrone is overplotted (black line). }
\label{fig:CMD_empiric}
\end{center}
\end{figure*}

The empirical isochrones provide a key information to compare and constrain the theoretical evolutionary models. In this study, we use the membership analysis of IC~4665 to report the empirical isochrone of a 30~Myr old open cluster (see Fig.~\ref{fig:CMD_empiric}). 

To obtain the empirical isochrones, we fitted a principal curve to the members in several apparent colour magnitude diagrams. Then, we manually shifted the principal curve to reach the lower edge of the distribution which is supposed to correspond to the single star zero age main sequence (ZAMS). In addition, we applied manual offsets where needed to better fit the lower edge of the cluster sequence. The empirical isochrones we provide are thus the lower envelope sequence of the members. This does not correspond to the principal curve which indicates the mean position of the sequence.

In Tables~\ref{tab:isoc_empiric_GDR2} and \ref{tab:isoc_empiric_DANCe} we give the apparent magnitudes for the IC~4665 empirical sequence in the $G, G_{\rm BP}, G_{\rm RP},i,Y,J,H,Ks$ over the dynamic range of our dataset. We reiterate that they are just an estimation of the ZAMS locus at the age of IC~4665 which could be used to compare with other clusters. For more quantitative analysis, we provide the full list of members and let the interested user decide the most convenient way to use them.

\subsection{Evolutionary models}
\label{subsec:evolutionary_models}

\begin{figure*}
\begin{center}
\includegraphics[height=0.94\textheight]{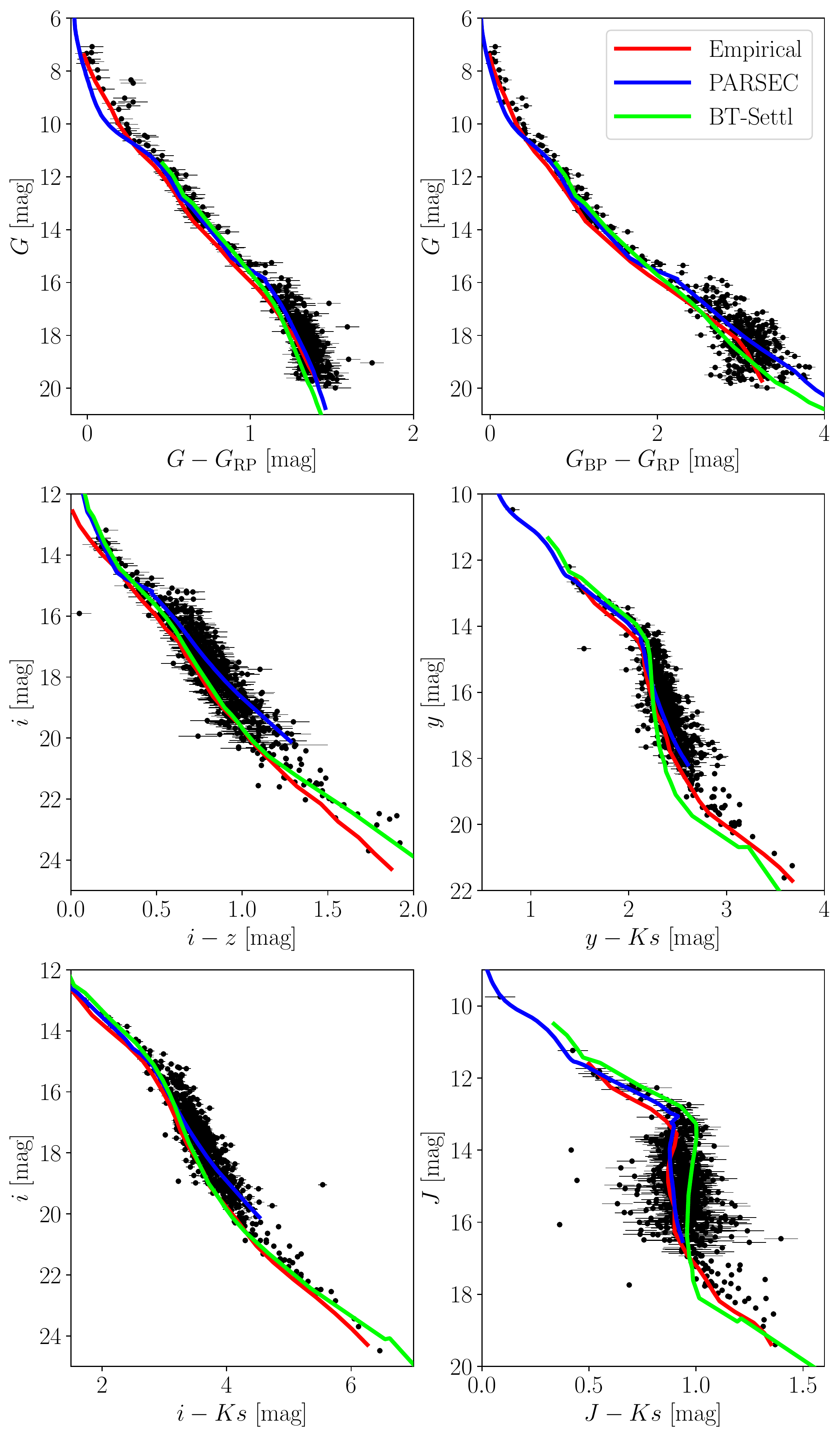}
\caption{Comparison of the observations of the IC~4665 sequence (black dots) and the empirical sequence (red lines) to the models of PARSEC+COLIBRI (blue line) and the BT-Settl (green line) for an age of 30~Myr in several colour magnitude diagrams.}
\label{fig:CMD_empiric_models}
\end{center}
\end{figure*}

In Figure~\ref{fig:CMD_empiric_models} we compare the observed sequence of IC~4665 and the empirical isochrones described in Sect.~\ref{subsec:empirical_isoc} to the 30~Myr models of PARSEC-COLIBRI\footnote{\url{http://stev.oapd.inaf.it/cgi-bin/cmd}} \citep{Marigo+17} and BT-Settl\footnote{\url{http://perso.ens-lyon.fr/france.allard/}} \citep{Allard14} in several colour magnitude diagrams. The distance modulus applied to the models uses the median parallax of the GDR2 members (2.81~mas). We did not include the effect of any systematic because here we only intend to do a qualitative comparison. We also corrected the models for the median extinction measured with \textit{Gaia} (A$_\textup{G}=0.62$~mag which corresponds to A$_\textup{V}=0.72$~mag using the A$_\textup{G}$/A$_\textup{V}=0.85926$ from the PARSEC website, see footnote 3).

As a general result, we see that the models show a major improvement with respect to previous versions, specially in the $y, J, H, Ks$ bands (see e.g. the comparison of the Pleiades by \citealt{Bouy+15}) even at such a young age. The brightest stars are only covered by the PARSEC isochrones while the faintest are only covered by the BT-Settl models. Between $i=11-15$~mag, both models agree fairly well between them and with the observations. However, the PARSEC models start to differ from the observations at $i>15$~mag, and in this magnitude range the BT-Settl models are believed to be more accurate. Despite the global improvement of the models in all the photometric bands, we still find a space for improvement in some of them, specially the ones involving the redder bands (see middle left and bottom left panels of Fig.~\ref{fig:CMD_empiric_models}). For this, low contaminated samples combined with accurate photometric measurements along a wide magnitude range are essential.

Regarding the \textit{Gaia} DR2 photometry, it is noticeable that the $G_{\rm BP}$ band shows a larger spread for magnitudes $>18$ mag. In the near-infrared, our measurements come mostly from 2MASS that has relatively large errors beyond 14~mag explaining the larger dispersion between $14<J<17$. Beyond, the measurements come from our own deeper images and both the uncertainties and the dispersion of the isochrone are significantly smaller.

\subsection{The absolute colour magnitude diagram}
\label{subsec:absCMD}

\begin{figure}
\begin{center}
\includegraphics[width = 0.45\textwidth]{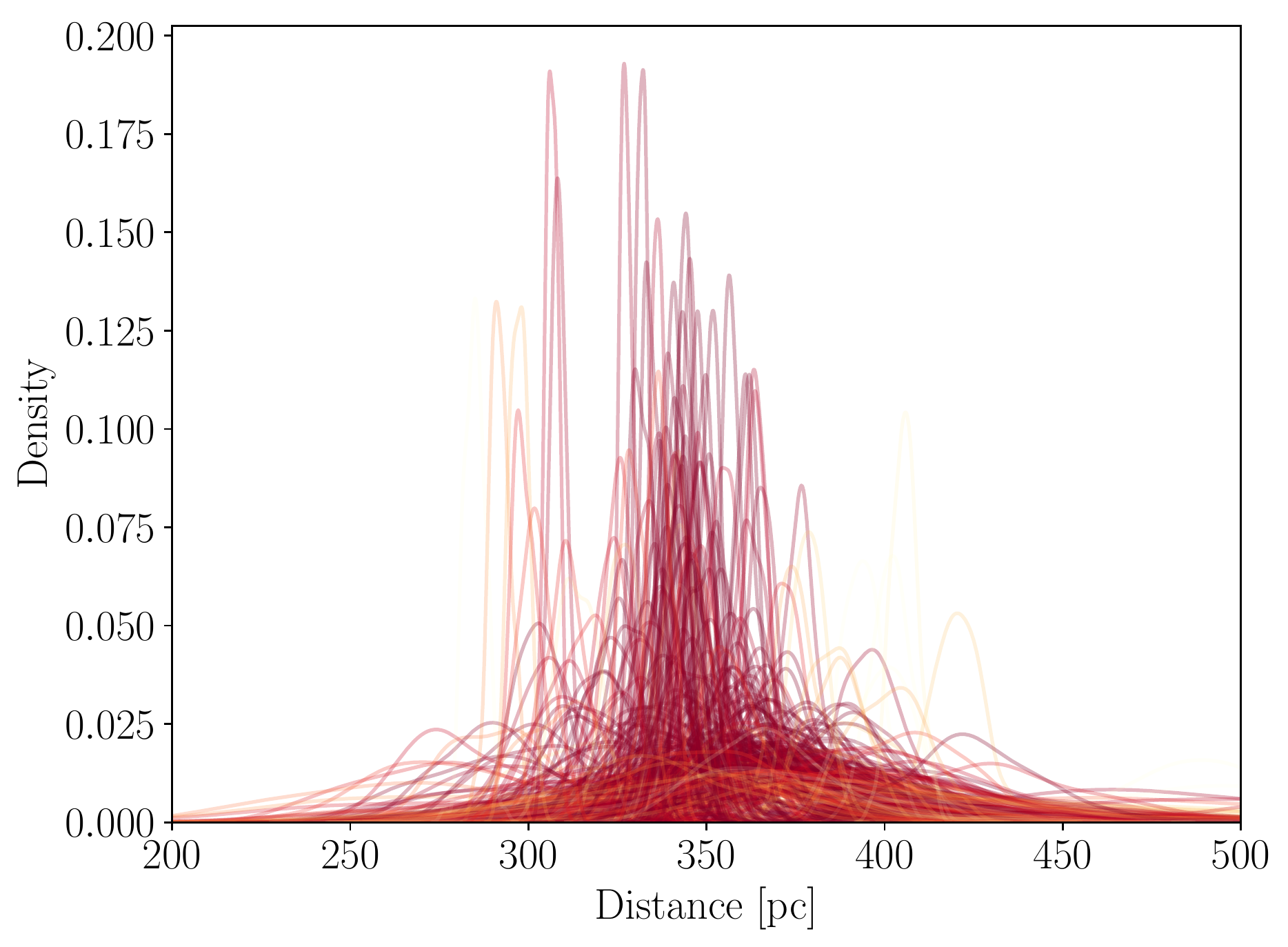}
\caption{Individual posterior distance distributions of the members of the GDR2 analysis computed with the \textit{Kalkayotl} code. }
\label{fig:distances}
\end{center}
\end{figure}

\begin{figure*}
\begin{center}
\includegraphics[width = 1\textwidth]{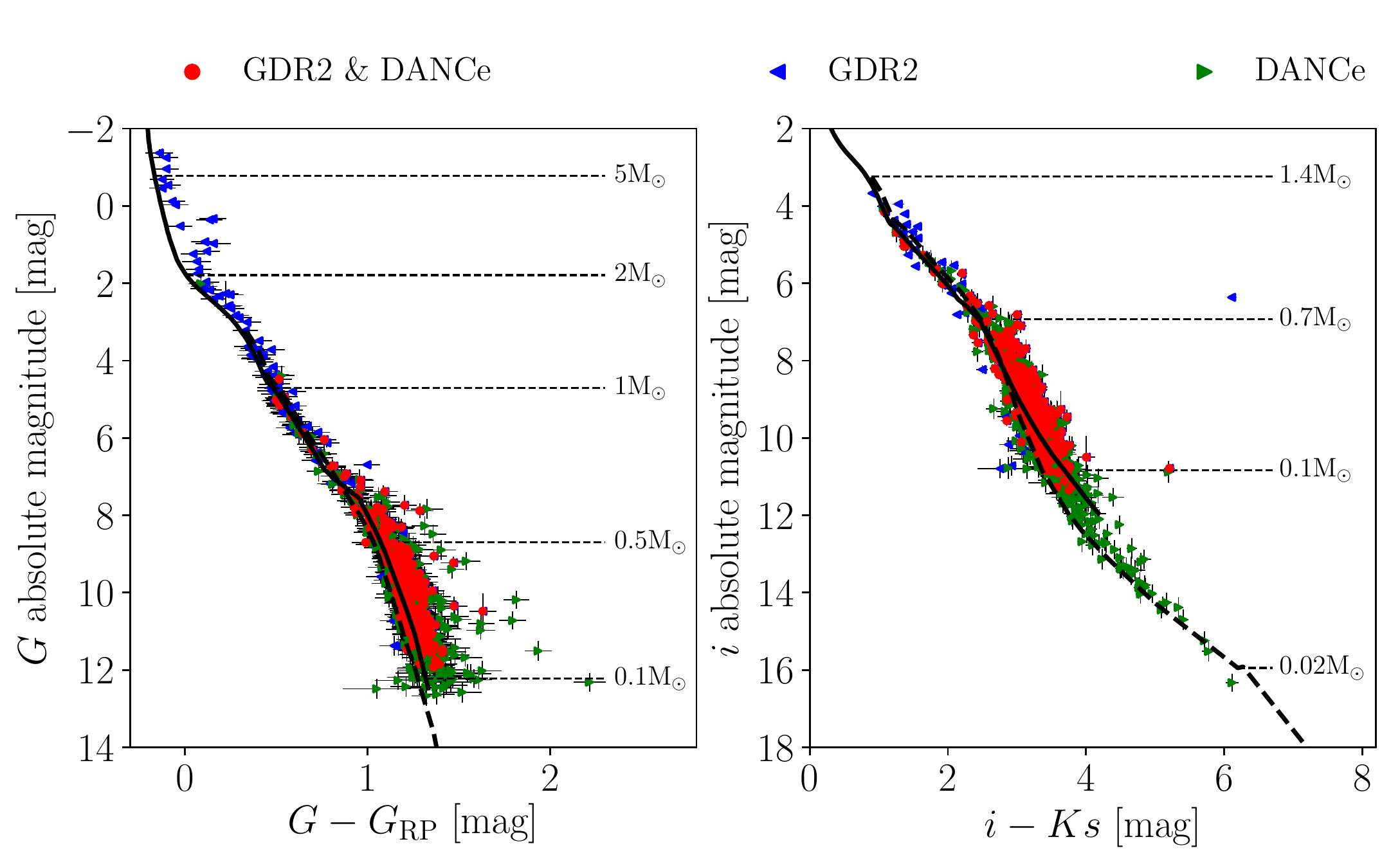}
\caption{(Absolute $G$, $G-G_{\rm RP}$) and (Absolute $i$, $i-Ks$) colour magnitude diagram (left and right, respectively) of the IC~4665 open cluster. The PARSEC+COLIBRI (solid line) and the BT-Settl (dashed line) isochrones of 30~Myr are overplotted.  The members are shape and colour coded according to their membership classification: red circles for members in GDR2 and DANCe analysis, blue left-pointing triangles for members only in GDR2 and green right-pointing triangles for members only in DANCe.}
\label{fig:CMD_models}
\end{center}
\end{figure*}

To build the absolute colour magnitude diagram, we first converted individual apparent magnitudes to absolute magnitudes. This transformation requires the distance and extinction of each source. Given the different origin of our two catalogues (\textit{Gaia} provides individual parallaxes and DANCe does not) we decided to follow two different approaches.

For the members obtained with the GDR2 membership analysis we used individual parallaxes to compute distances. We used the \textit{Kalkayotl}\footnote{\url{https://github.com/olivares-j/kalkayotl}} code as in \cite{Olivares+19} which performs a Bayesian probabilistic inference to compute posterior probability distributions for the distance of each member. We chose a Cauchy prior which is the recommended for clusters by the manual. The location of the prior was set to 350~pc (the approximate distance of the cluster), and the scale to 100~pc (in order to have a loose prior). In Figure~\ref{fig:distances} we show the individual posterior distance distribution of each of the GDR2 members. The median distance is 351~pc and the standard deviation is 55~pc. 

We estimated the extinction in two independent ways. First, we used the \textit{Gaia} extinction estimate (\texttt{a\_g\_val}) for the 88 members which have it available. This value is recommended not to be used individually but statistically on a set of stars \citep{Andrae+18}. Thus, we compute the median A$_\textup{V}=0.72$~mag and the standard deviation 0.38~mag. Aside, we inferred the individual absorption of each source using a Bayesian model as in \citealt{Olivares+19} (see Sect.~\ref{subsec:PDMF}). In this case, we also compute the median of all the individual maximum a posteriori probabilities (MAP) which is A$_\textup{V}=0.66$~mag, compatible with the extinctions from \textit{Gaia}. 

To compute the absolute magnitude we sampled the apparent magnitude with a Gaussian centred at the observed magnitude and a standard deviation equal to the uncertainty. Then, each sample was converted to absolute magnitude by sampling the posterior distance distribution obtained with \textit{Kalkayotl}. We added the median absorption of the cluster to each member.

To compute absolute magnitudes for the DANCe members we followed a similar approach. The only difference is that instead of sampling the distance from the individual posterior distributions, we sampled the distance from the cluster distribution obtained with all the GDR2 members. 

In Figure~\ref{fig:CMD_models} we show the absolute colour magnitude diagram of IC~4665 where we have overplotted the PARSEC and the BT-Settl models. Thanks to the precision of the \textit{Gaia} DR2 parallaxes, we find that the isochrone has broadened little with respect the the one observed in the apparent colour magnitude diagrams. In addition, we have included a mass scale. We have candidate members down to masses of $\sim0.02$~M$_\sun$, well within the sub-stellar regime. We see that the PARSEC models start to differ for masses $<0.7$~M$_\sun$ and in this low-mass regime the BT-Settl models represent better the observations. For this reason, to convert magnitudes to masses, we use the PARSEC models for the high mass stars and the BT-Settl models for low-mass stars (see Sect.~\ref{subsec:PDMF}).

\section{From the apparent magnitude distribution to the present-day system mass function}
\label{sec:mass-func}

\subsection{The apparent magnitude distribution}
\begin{figure*}
\begin{center}
\includegraphics[width = 1\textwidth]{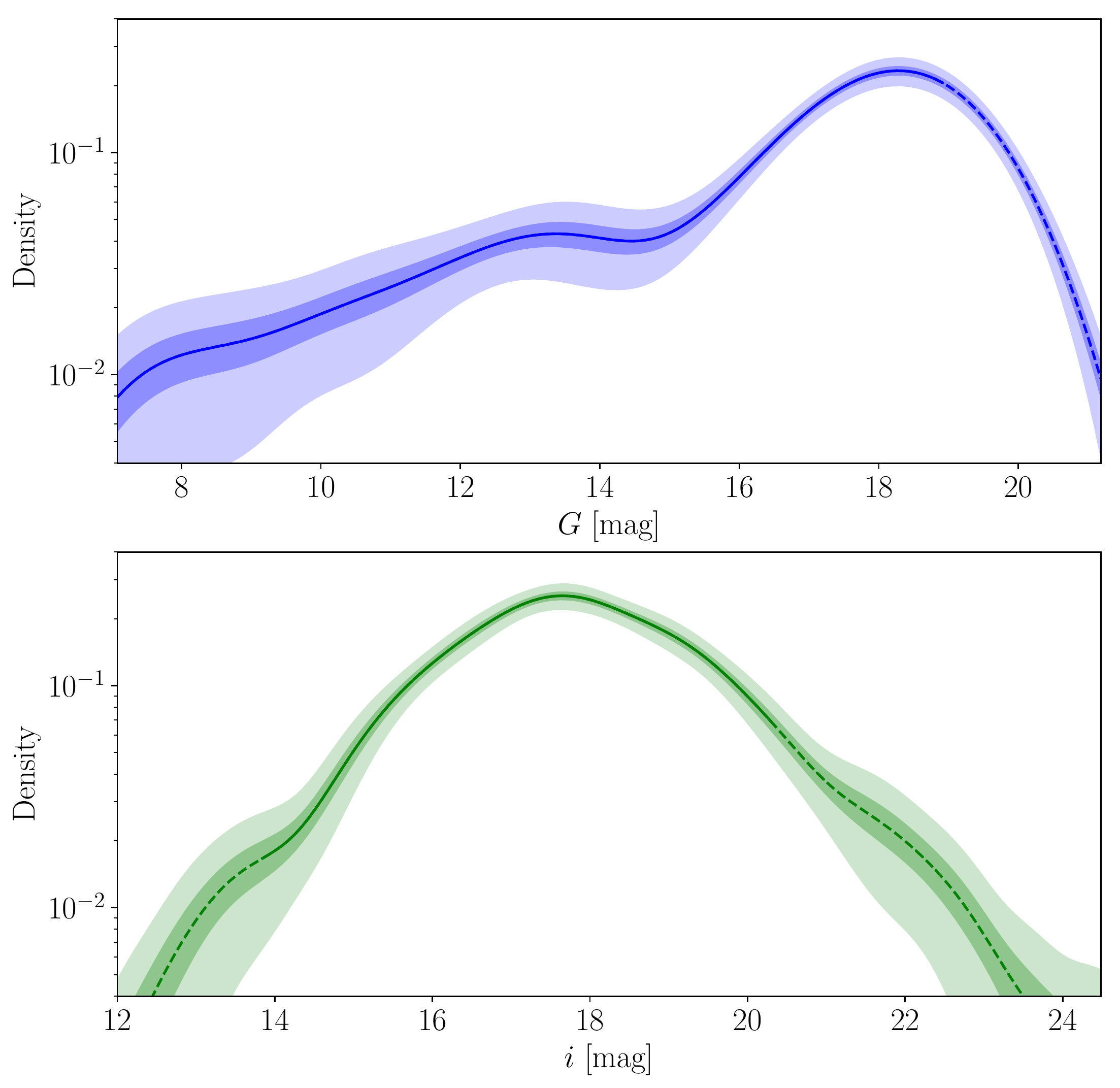}
\caption{Top: $G$ magnitude distribution of IC~4665 with the members found with the GDR2 catalogue. Bottom: $i$ magnitude distribution of IC~4665 with the members found with the DANCe catalogue. In both cases, the shaded regions indicate the 1 and 3$\sigma$ uncertainties estimated from bootstrap (1$\sigma$ dark and 3$\sigma$ faint). The dashed lines indicate the region of incompleteness in each catalogue. }
\label{fig:mag_dist}
\end{center}
\end{figure*}

The apparent magnitude distribution is a direct measurement of the number of sources observed at different brightnesses. The importance of this function lies in the fact that it does not depend on evolutionary models nor on distance estimates and thus, its validity does not expire (unless selection problems are present). The magnitude distribution of IC~4665 was obtained applying a Gaussian kernel density estimation (KDE) independently to the GDR2 and the DANCe members. The two samples were treated independently because of the different validity range of each catalogue. To estimate the optimal bandwidth of the KDE we considered the Scott's rule \citep{Scott+92} and the Silverman's rule \citep{Silverman+86}, which gave similar results. The optimal bandwidth of the KDE was 0.3~mag (both in the GDR2 and DANCe members), and the uncertainties were estimated by means of a bootstrap with 100 repetitions. We have estimated the effect of contamination and completeness as a function of the magnitude using synthetic data as  in \citet{Olivares+19}. Given that the contamination rate estimated this way is less than 15\%, we realised that when we correct for these two effects the magnitude distribution we obtain is compatible with the original one within the uncertainties. For this reason, we decided to work with the magnitude distribution we obtain directly from the observations.  In Fig.~\ref{fig:mag_dist} we show the magnitude distribution of IC~4665 in the $G$ band for the GDR2 members and in the $i$ band for the DANCe members. These functions are available in Tables~\ref{tab:G_dist} and \ref{tab:i_dist}, respectively. 

At $G\sim14.5$~mag there is a flattening of the apparent magnitude distribution which corresponds to the Wielen dip \citep{Wielen+83}. This feature has been reported in other open clusters such as the Pleiades \citep{Lee+95, Belikov+98}, Praesepe and Hyades \citep{Lee+97}, NGC 2516 \citep{Jeffries+01}, NGC 2547 \citep{Naylor+02}, and Ruprecht 147 \citep{Olivares+19}. \cite{Kroupa+90} explained this dip as the result of a change in the opacities in the corresponding mass range.

The apparent magnitude distribution peaks at $G=18.2$~mag for the GDR2 members and at $i=17.6$~mag for the DANCe members which in both cases correspond to $0.2$~M$_\sun$, according to the PARSEC and BT-Settl models and assuming an age of 30~Myr. This result is in agreement with what \cite{Bouy+15} found in the Pleiades. 

A change of slope seems to happen around $i\sim21$~mag, which could indicate that different formation mechanisms are at work for ultracool objects in this mass range. This change of slope is nevertheless very close to our estimated limit of completeness and not statistically significant yet.

\subsection{The present-day system mass function}
\label{subsec:PDMF}
\begin{figure*}
\begin{center}
\includegraphics[width = 1\textwidth]{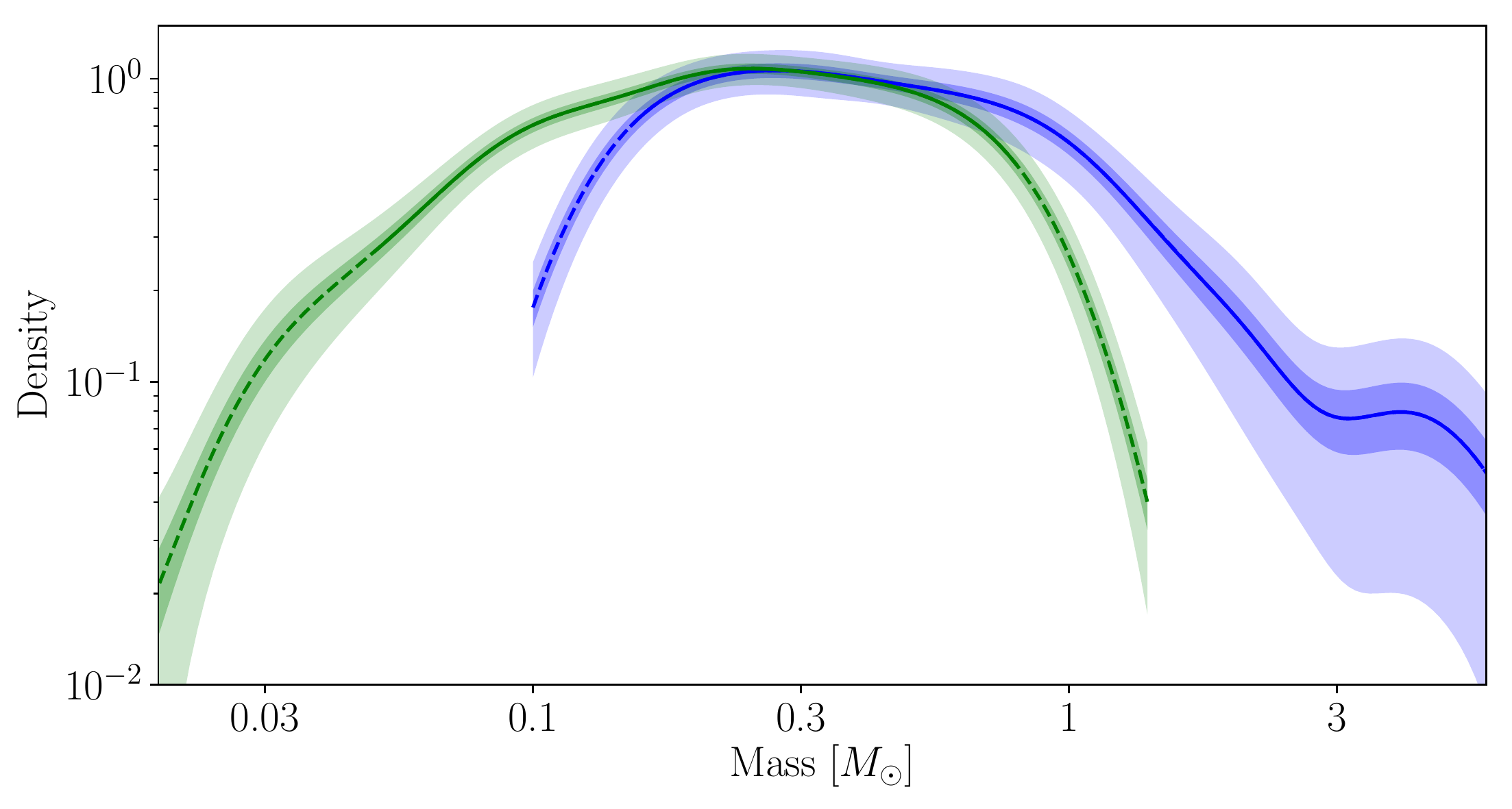}
\caption{Present-day system mass function of IC~4665 obtained from the GDR2 (blue) and DANCe (green) members. The shaded regions indicate the uncertainty estimated from bootstrap (1$\sigma$ dark and 3$\sigma$ faint) and the dashed lines the regions of incompleteness. }
\label{fig:mass-func}
\end{center}
\end{figure*}

We estimated the mass of each source using the \textit{Sakam}\footnote{\url{https://github.com/olivares-j/Sakam}} code \citep{Olivares+19}. This algorithm infers the posterior distribution of the mass together with the A$_\textup{V}$ extinction, given the absolute photometry of each source (computed in Sect.~\ref{subsec:absCMD}) and a theoretical evolutionary model. The model does not include effects in the variability of the source due to binarity, activity or others. These effects are eventually included in the extinction estimate, enlarging its uncertainties. We used the PARSEC model to infer masses for the GDR2 members and the BT-Settl model for the DANCe members.

To compute the present-day system mass function, we took samples of the a posteriori distribution inferred with \textit{Sakam} and applied a Gaussian KDE with a bandwith of 0.3 (in $\log_{10}m$ where $m$ is the mass, and the bandwidth of the KDE was estimated from the Scott's and Silverman's rule). We estimated the uncertainties from 100 bootstrap repetitions and reported the 1 and 3$\sigma$ confidence levels. The completeness limits were propagated from the catalogue completeness in apparent magnitude (see Table~\ref{tab:missing_phot}). The mass function obtained with the DANCe analysis was renormalised so that the mass distribution functions had the same area in the region where both studies are complete, i.e. 0.15--0.8~M$_\sun$.

Figure~\ref{fig:mass-func} shows the present-day system mass function of IC~4665 for the GDR2 (blue) and DANCe (green) members. We see that the two functions overlap reasonably well. There are some deviations (even inside the complete range) but they are smaller than $3 \sigma$. The robustness of our methodology, specially in the error propagation, results in a mass function with an accuracy significantly better than in the past \citep[i.e.][]{deWit+06, Lodieu+11}. 

A number of noticeable details are present in the mass function. At 3~M$_\sun$ we observe a feature which has not been reported in the literature before. It is not clear whether this is a real feature of the mass function or an artefact. Several sources of error could be responsible for that, in particular:
\begin{itemize}
\item the uncertainties or errors of the transformation from apparent magnitudes to masses, since it is not observed in the magnitude distribution (Fig.~\ref{fig:mag_dist}),
\item multiplicity: the \textit{Gaia} DR2 catalogue excluded a number of binary stars. Since massive stars are more often in multiple systems than their lower mass counterparts \citep[e.g.][]{Lada06}, we might be missing a larger fraction of massive members because of multiplicity, 
\item variability: massive stars can also display photometric variability, which is not included in our algorithm to determine individual masses. Slowly Pulsating variable stars appear at three solar masses and beyond. However, these are small amplitude variables (0.1 in $V$) and should not have a major impact in our selection.
\end{itemize}
Additionally, and as mentioned in Section~\ref{sec:gaiadr2}, we assumed that the GDR2 catalogue is complete for $G>7$~mag ($\sim5$~M$_\sun$) but a few sources between $7<G<12$~mag ($5\lesssim m\lesssim1.6$~M$_\sun$) could be missing \citep{GaiaColBrown+18}. Nevertheless, this would only increase the number of members in this range.

The Wielen dip reported in the magnitude distribution (Fig.~\ref{fig:mag_dist}) would be expected around 0.75~M$_\sun$ in the mass distribution but is not observed. If confirmed, this result would support the hypothesis of \cite{Kroupa+90} explaining this feature by a change in opacity rather than a change in the mass function. We nevertheless note that the Wielen dip may have been masked by the KDE bandwidth. \citet{Olivares+19} indeed reported a Wielen dip in their mass function between 0.6--0.8~M$_\sun$ with a typical scale of $\Delta\log_{10}m\sim0.13$, smaller than our bandwidth of 0.3 (in $\log_{10}m$).

The function is rather flat between 0.1 and 1~M$_\sun$ having a maximum at 0.28~M$_\sun$. For masses $<0.1$~M$_\sun$ the distribution drops. The change of slope at the very low mass end mentioned above is not visible in the mass function.

The highest mass object has a MAP estimate of 6.2~M$_\sun$ and the lowest mass object has a MAP estimate of $13$~M$_\textup{J}$ according to the PARSEC and BT-Settl models respectively, and assuming an age of 30~Myr. To compute the \textit{brown dwarf-to-star ratio} we sampled the posterior mass distribution of each member. Then, we used these samples to compute the ratio of brown dwarfs and stars within the completeness region of our sample (6--0.05~M$_\sun$) and using a mass threshold of $0.08$~M$_\sun$. We did a bootstrap over all the members with 100 repetitions and, we obtained a median ratio of $0.067\pm0.005$. This value is lower to what has been seen in other nearby young clusters as in IC~348 and Taurus \citep[][and references therein]{Scholz+12}. However these studies are complete down to lower masses ($\sim0.02$M$_\sun$).

\subsection{Comparison to other clusters and theoretical models}

\begin{figure*}
\begin{center}
\includegraphics[width = 1\textwidth]{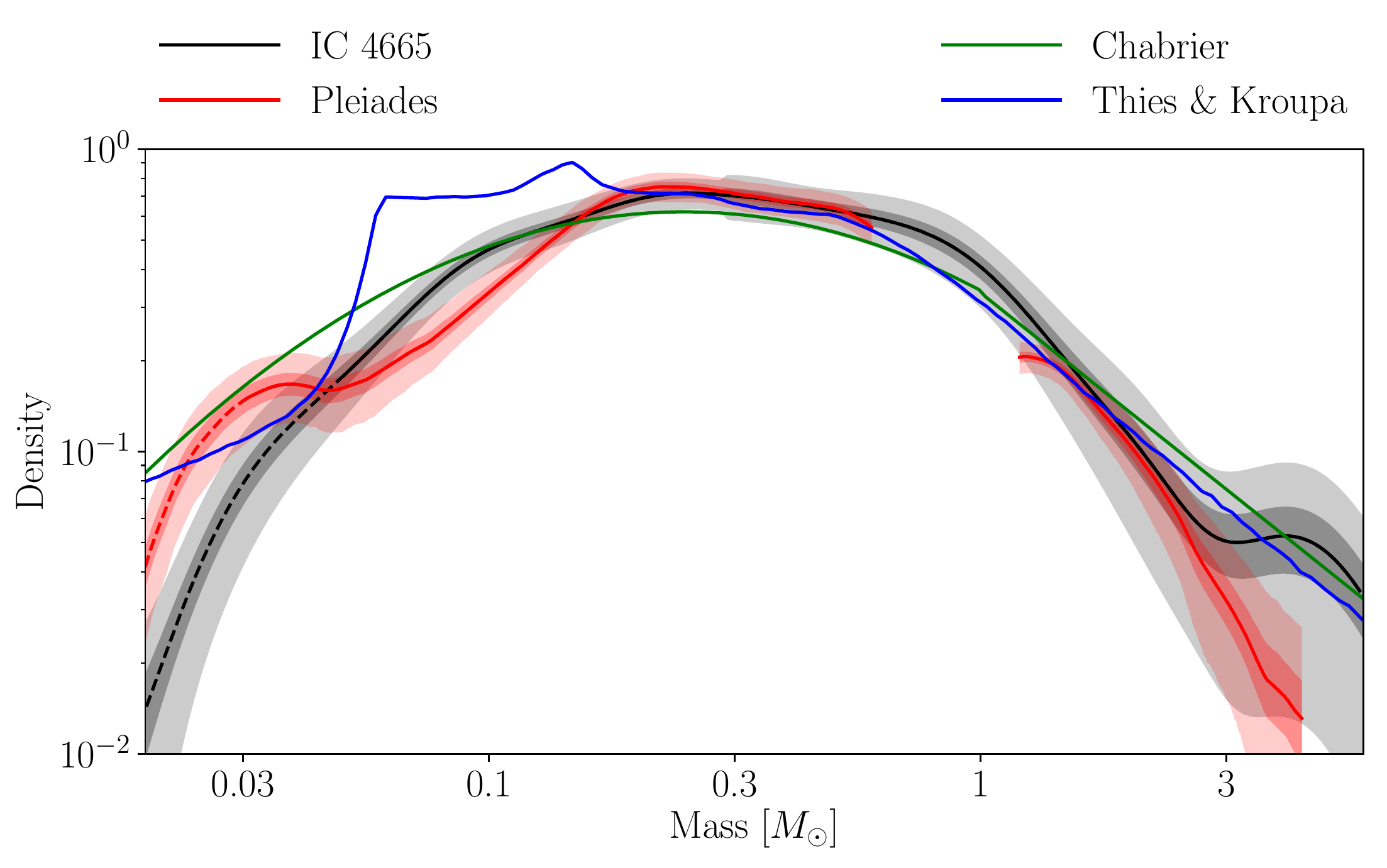}
\caption{Present-day system mass function of IC~4665 (black) and the Pleiades (red). The shaded regions indicate the uncertainties estimated from bootstrap (1$\sigma$ dark and 3$\sigma$ faint). Over-plotted are the models from \citealt{Chabrier05} (green) and \citealt{Thies+15} (blue). }
\label{fig:mass-func-clust}
\end{center}
\end{figure*}

In Figure~\ref{fig:mass-func-clust}, we compare the PDMF we obtained for the 30~Myr open cluster IC~4665 to the one from the Pleiades \citep[120 Myr;][]{Bouy+15}. To facilitate the comparison, we normalised the mass function of IC~4665 over the whole mass range where it is complete. Then, we normalised the Pleiades mass function so that it had the same area between 0.05--0.6~M$_\sun$, range where both functions are complete. 
We see that in general both functions match fairly well within the uncertainties and the main differences are observed at the extremes of the distributions. For the high mass domain, IC~4665 has more massive stars than the Pleiades. In this range the number of members is quite small (only 12 objects have masses >3~M$_\sun$ in IC~4665) leading to rather large statistical uncertainties. In addition, multiplicity (more frequent among high mass stars) and variability affect the luminosities and might contribute to the differences observed.   
Regarding the low-mass regime, we see that both functions are compatible (within $3\sigma$ uncertainties) down to the IC~4665 completeness limit ($\sim$0.05~M$_\sun$). Nonetheless, we observe that between 0.046 and 0.16~M$_\sun$ the mass function of IC~4665 might display a slight over-density with respect to the Pleiades at the 1$\sigma$ level only. For masses lower than 0.05~M$_\sun$, the Pleiades mass function exhibits a change of slope which the authors related to a different mechanism of star formation for this regime of masses. In the case of IC~4665, we do not detect this change of slope but this could be because it is beyond the completeness limit of the catalogue. 

In Figure~\ref{fig:mass-func-clust} we have overplotted the sytem IMF of two models, namely \citet{Chabrier05} and \cite{Thies+15}, normalised in the same mass range as the mass function of IC~4665. In the high mass regime ($>1$~M$_\sun$), both models assume
a power law IMF with Salpeter slope which is compatible within the uncertainties with the empirical mass function of IC~4665.

For intermediate and low masses, we see that the mass function of IC~4665 is compatible with the model of \citet{Chabrier05} between 0.1--1~M$_\sun$. For lower masses, the model predicts too many stars compared to our results. The model of \citet{Thies+15} is compatible with the empirical mass function between 0.2--1~M$_\sun$ but between 0.05--0.2~M$_\sun$ it also predicts too many stars. Below 0.05~M$_\sun$ the model approaches the empirical mass function and beyond this limit our survey is not complete.

\section{Projected spatial distribution}
\label{sec:spatial_dist}

\begin{table*}
    \centering
    \caption{Median parameters for the spherically symmetric distributed models. We have assumed the median distance of the cluster (350~pc) as the distance estimate.}
    \begin{tabular}{l|ccccccc}
        \hline
		\hline
		Model & $\alpha_c$ & $\delta_c$ & $r_c$ & $\gamma$ & $\alpha$ & $\beta$ & $r_t$ \\ 
		      & [\degr]    & [\degr]    & [pc]  &          &          &         & [pc] \\ 
		\hline
		EFF & $266.573^{+0.070}_{-0.069}$ & $5.439^{+0.092}_{-0.101}$ & $2.27^{+1.22}_{-0.36}$ & $2.129^{+0.314}_{-0.090}$ & -- & -- & -- \\ 
		GDP & $266.580^{+0.051}_{-0.067}$ & $5.439^{+0.111}_{-0.056}$ & $2.10^{+0.52}_{-0.95}$ & $0.18^{+0.32}_{-0.13}$ & $0.20^{+0.59}_{-0.15}$ & $1.81^{+0.16}_{-0.32}$ & -- \\ 
		GKing & $266.585^{+0.049}_{-0.061}$ & $5.432^{+0.102}_{-0.061}$ & $1.87^{+0.44}_{-0.97}$ & -- & $0.21^{+0.64}_{-0.15}$ & $1.67^{+0.18}_{-0.41}$ & $92^{+306}_{-59}$ \\ 
		King & $266.573^{+0.073}_{-0.069}$ & $5.442^{+0.090}_{-0.105}$ & $2.12^{+0.91}_{-0.34}$ & -- & -- & -- & $190^{+620}_{-140}$ \\ 
		OGKing & $266.576^{+0.067}_{-0.066}$ & $5.452^{+0.098}_{-0.086}$ & $1.55^{+0.47}_{-0.44}$ & -- & -- & -- & $55^{+50}_{-16}$ \\ 
		RGDP & $266.583^{+0.053}_{-0.070}$ & $5.433^{+0.112}_{-0.062}$ & $1.98^{+0.43}_{-0.93}$ & -- & $0.22^{+0.58}_{-0.15}$ & $1.83^{+0.18}_{-0.25}$ & -- \\ 
		\hline
    \end{tabular}
\label{tab:spatial_distribution}
\end{table*}

\begin{figure*}
\begin{center}
\includegraphics[width = 0.95\textwidth]{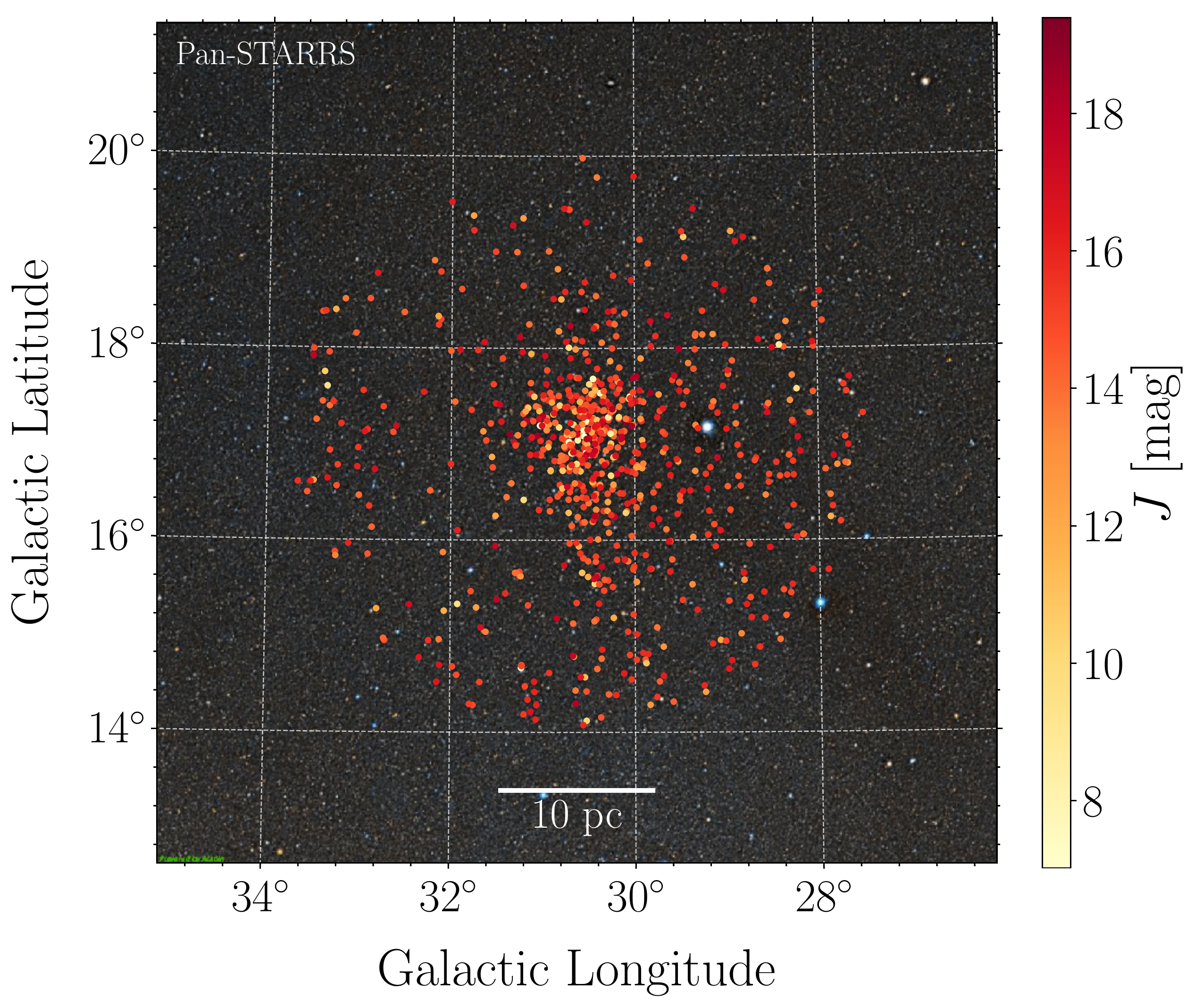}
\caption{Spatial distribution of the members of IC~4665. The members are colour coded according to their magnitude in $J$ band. Background image credit: Pan-STARRS. }
\label{fig:spatial-distribution}
\end{center}
\end{figure*}

The spatial distribution of open clusters provides relevant information on the formation and early evolution of these systems. In Figure~\ref{fig:spatial-distribution} we show the spatial distribution of the members of IC~4665 in galactic coordinates. At first glance we can intuit some structures which depart from a pure spherical symmetry (e.g. the cluster seems elongated towards the Galactic south). In this section, we apply a statistical treatment to quantitatively asses the probability that the structures we might see are significant.

We follow the same approach as \cite{Olivares+18} and we fit a series of parametric models to the projected spatial distribution of the cluster (i.e. in the plane of the sky). The algorithm they used, \textit{PyAspidistra}\footnote{https://github.com/olivares-j/PyAspidistra}, computes the Bayesian evidence of each model and the posterior distribution of the parameters which characterise the model. Eventually, they compare the Bayesian evidence of each pair of models by means of the Bayes Factor. Here, we consider the same set of models as those authors: the \citealt{Elson+87} (hereafter EFF), the Generalized Density Profile \citep[hereafter GDP, also known as Nukker,][]{Kupper+10}, the \citealt{King62} (herafter King), the Generalised King \citep[hereafter GKing, ][]{Olivares+18}, the Optimum Generalized King \citep[herafter OGKing, ][]{Olivares+18}, and the Restricted Generalized Density Profile \citep[herafter RGDP, ][]{Olivares+18}. For each model, the \textit{PyAspidistra} code has the option to infer the coordinates of the cluster centre, its ellipticity, and mass segregation. 

Using the equatorial coordinates (J2000), the median distance of the cluster (350~pc), and the $J$ band we ran the \textit{PyAspidistra} code and obtained the Bayesian evidence for each model and the Bayes Factor for each pair of models. The RGDP model is the one that shows the largest evidence in all the family models considered (spherical, elliptical and segregated). The family of models with largest evidence are the spherical models, as it is expected for such a young open cluster. However, the strength of this evidence is weak according to the criterion from \citet{Jeffreys61} so these results should be taken with care and we can not definitely discard the possibility of ellipticity or mass segregation. In addition, our results could be biased due to the contamination in the members and the size and shape of our initial catalogues as we discuss below. 

The median parameters of each spherical model are reported in Table~\ref{tab:spatial_distribution}. The parameters $\alpha_c$ and $\delta_c$ correspond to the central coordinates of the cluster in RA and Dec, respectively. The core radius ($r_c$) is the unit scale of the density profile, therefore it differs for each model. The $\alpha, \delta$, and $\gamma$ parameters correspond to the exponents of the different models. The tidal radius ($r_t$) is only defined for the family of King's models. We refer the interested reader to \citet{Olivares+18} for a detailed discussion of these parameters. We see that the centre of the cluster is well determined by all the models at RA = 266.6\degr, Dec = 5.4\degr. All the models also agree to a core radius of $\sim2$~pc and the small dispersion is expected since this parameter has a different interpretation in each model. Only the family of King models are defined in terms of a tidal radius. The median values of the tidal radius reported in Table~\ref{tab:spatial_distribution} vary from one model to another and have extremely large uncertainties. The King model with larger evidence is the OGKing model which predicts a $r_t=55$~pc, 3 times larger than the radius analysed in this study (18~pc). This is probably the main reason why we fail to well constrain this parameter (see also the discussion on the main caveats at the end of the section). However, we note that up to the present, this study of IC~4665 is the one with largest radius. Our estimate of the tidal radius is remarkably larger than previous values (e.g. \citealt{deWit+06} reported a tidal radius of 1\degr corresponding to $\sim6$~pc at the distance of the cluster; however, this value results from a highly contaminated sample).  

Here we enumerate some of the caveats and limitations of our study of the spatial distribution.
\begin{itemize}
    \item Our members come from a catalogue which was circularly selected and the ends of the catalogue can clearly be seen in Fig.~\ref{fig:spatial-distribution}. This can bias our results to favour a circular over an elliptic model.
    \item The spatial coverage of the DANCe catalogue (see Sect.~\ref{subsec:DANCe_dataset} and \citealt{Bouy+13}) implies that the faintest members are more likely found in the centre. This region is where we have not only the highest number of images, but also the deepest ones and the ones with the longest time baseline. As a consequence, the proper motions also have in general better precision in this area, which in turn has an influence on the membership probabilities. This can have an impact both in the study of the shape (circular or elliptical) and the study on the segregation.
    \item The tidal radius and the contamination rate are degenerate: a large contamination rate increases the density of members and the models need a larger tidal radius to explain the observations. This is specially critical at the outskirts, where we believe we might have a larger contamination. As we have discussed in Sect.~\ref{subsec:final_memb_list} and Sect.~\ref{sec:analysis-members}, the majority of our contaminants come from the DANCe catalogue.
    \item Our members come from a catalogue which was truncated to a radius of $\sim$18~pc, similar to the expected tidal radius. It is highly difficult to estimate the tidal radius without having information beyond it. All this difficulties are reflected on the Bayesian evidence of each model: the family of King's models have lower evidences. The truncation in radius could also bias the study on the elipticity and segregation of the cluster.
\end{itemize}

\section{Conclusions}
\label{sec:conclusions}
We presented an exhaustive study of the properties of the IC~4665 open cluster. We combined the recent \textit{Gaia} DR2 data with the deep, on ground observations of the COSMIC DANCe project to search for members. We used the same methodology as \citet{Olivares+19} to derive Bayesian posterior probabilities for all the sources and found 819 members, 50\% of which are new members. Our members have magnitudes in the range $7<J<19.4$, which correspond to masses between 6.2~M$_\sun$--13~M$_\textup{J}$, according to the PARSEC and BT-Settl evolutionary models, and assuming an age of 30~Myr. Using this sample, we provided the empirical isochrones of the cluster, an estimate for the distance, the magnitude distribution, the present-day system mass function with unprecedented accuracy for this cluster, and a study of the spatial distribution. 

Comparing our members with previous studies in the literature, we found that most of the previous studies were based on highly contaminated (up to 80\%) or incomplete samples. The low motion of this cluster with respect to the field ($<10$~mas~yr$^{-1}$) complicates the membership analysis. For this reason we found a larger contamination rate in this study compared to others which use the same methodology applied to clusters with larger proper motions \citep{Sarro+14, Olivares+19}. Using synthetic data, we estimated a CR of 10\% for GDR2 and a 13\% for DANCe. Comparing the two studies, we estimate that the DANCe contamination reaches up to 30\% in the region of completeness. The main reason of this underestimated CR in the DANCe study is the lack of parallaxes. Anyway, this study provides the most accurate membership analysis up to date by far and thus, offers the possibility to revisit other fundamental parameters such as the age.

We found that the PDMF of IC~4665 in the intermediate mass range (0.1--1~M$_\sun$) is comparable to that of the Pleiades \citep{Bouy+15} and to models of the IMF \citep{Chabrier05, Thies+15}. For higher masses, the observations have a steeper slope than that of the models (Salpeter slope). In addition, for the case of IC~4665, we find a plateau at 3~M$_\sun$ which if further confirmed would represent a new feature of the mass function. In the mass range 0.05--0.2~M$_\sun$ the models predict too many low-mass stars. For masses lower than 0.05~M$_\sun$ the Pleiades have more members than in IC~4665, but at this mass regime our study is not complete so we might be missing members.

Combining our comprehensive census of the cluster with the \textit{Gaia} DR2 parallaxes we estimated the distance of the cluster to be of 350~pc, this value being similar to what other studies recently derived \citep{GaiaColBabusiaux+18}. We found that the best surface density profile for IC~4665 is the RGDP model with a core radius of 2~pc, and sperical symmetry. However, we can not definitely discard the possibility of ellipticity nor mass segregation. In the future, we aim to include a study on the velocity distribution which would allow us to characterise the kinematic and dynamic state of the cluster in the 6D space phase. 

\begin{acknowledgements}
We thank the referee for his/her thorough review and highly appreciate the comments and suggestions, which significantly contributed to improving the quality of the publication.
We are grateful to Ingo Thies for sharing with us his models of the Initial Mass Function, to France Allard and Isabelle Barraffe for providing us with the latest version of the BT-Settl models, and to Paul Price for his help with the LSST/HSC pipeline.
This research has received funding from the European Research Council (ERC) under the European Union’s Horizon 2020 research and innovation programme (grant agreement No 682903, P.I. H. Bouy), and from the French State in the framework of the ”Investments for the future” Program, IdEx Bordeaux, reference ANR-10-IDEX-03-02 . This Project has been funded by the Spanish State Research Agency (AEI) Project No.ESP2017-87676-C5-1-R and No. MDM-2017-0737 Unidad de Excelencia “María de Maeztu”- Centro de Astrobiología (INTA-CSIC). M.T. is supported by MEXT/JSPS KAKENHI grant Nos. 18H05442, 15H02063, and 22000005.
Based on observations collected at the European Southern Observatory under ESO programmes 60.A-9120(A), 60.A-9121(A), 60.A-9122(A), 68.C-0311(A), 69.A-9014(A), 69.C-0034(A), 69.C-0260(A), 69.C-0398(B), 69.C-0426(C), 69.D-0582(A), 60.A-9038(A).
This research draws upon data distributed by the NOAO Science Archive. NOAO is operated by the Association of Universities for Research in Astronomy (AURA) under cooperative agreement with the National Science Foundation. This publication makes use of data products from the Two Micron All Sky Survey, which is a joint project of the University of Massachusetts and the Infrared Processing and Analysis Center/California Institute of Technology, funded by the National Aeronautics and Space Administration and the National Science Foundation.  
This work has made use of data from the European Space Agency (ESA) mission {\it Gaia} (\url{http://www.cosmos.esa.int/gaia}), processed by the {\it Gaia} Data Processing and Analysis Consortium (DPAC, \url{http://www.cosmos.esa.int/web/gaia/dpac/consortium}). Funding for the DPAC has been provided by national institutions, in particular the institutions participating in the {\it Gaia} Multilateral Agreement.
This research has made use of the VizieR and Aladin images and catalogue access tools and of the SIMBAD database, operated at the CDS, Strasbourg, France. This publication makes use of data products from the Wide-field Infrared Survey Explorer, which is a joint project of the University of California, Los Angeles, and the Jet Propulsion Laboratory/California Institute of Technology, funded by the National Aeronautics and Space Administration. This research has made use of GNU Parallel \citep{Tange2011a}, Astropy \citep{astropy:2013}, Topcat \citep{Taylor05}, STILTS \citep{Taylor06}.
\end{acknowledgements}


\begin{appendix}

\section{Queries}

\subsection{Gaia DR2}
\label{app:query_GDR2}
\begin{verbatim}
SELECT 
 "I/345/gaia2".ra\_epoch2000, 
 "I/345/gaia2".dec_epoch2000,  
 "I/345/gaia2".ra_epoch2000\_error, 
 "I/345/gaia2".dec\_epoch2000_error,  
 "I/345/gaia2".parallax,  
 "I/345/gaia2".parallax_error,  
 "I/345/gaia2".pmra,  
 "I/345/gaia2".pmra\_error, 
 "I/345/gaia2".pmdec,  
 "I/345/gaia2".pmdec\_error, 
 "I/345/gaia2".phot\_g\_mean\_mag, 
 "I/345/gaia2".phot_g_mean_mag_error, 
 "I/345/gaia2".phot_bp_mean_mag, 
 "I/345/gaia2".phot_bp_mean_mag_error, 
 "I/345/gaia2".phot_rp_mean_mag,  
 "I/345/gaia2".phot_rp_mean_mag_error, 
 "I/345/gaia2".radial_velocity,  
 "I/345/gaia2".radial_velocity_error,  
 "I/345/gaia2".teff_val,  
 "I/345/gaia2".a_g_val,  
 "I/345/gaia2".lum_val,  
 "I/345/gaia2".a_g_percentile_lower,  
 "I/345/gaia2".a_g_percentile_upper,  
 "I/345/gaia2".dec_parallax_corr, 
 "I/345/gaia2".dec_pmdec_corr, 
 "I/345/gaia2".dec_pmra_corr, 
 "I/345/gaia2".lum_percentile_lower,  
 "I/345/gaia2".lum_percentile_upper,  
 "I/345/gaia2".parallax_pmdec_corr,  
 "I/345/gaia2".parallax_pmra_corr,  
 "I/345/gaia2".pmra_pmdec_corr,  
 "I/345/gaia2".ra_parallax_corr, 
 "I/345/gaia2".ra_pmdec_corr, 
 "I/345/gaia2".ra_pmra_corr, 
 "I/345/gaia2".teff_percentile_lower, 
 "I/345/gaia2".teff_percentile_upper
FROM "I/345/gaia2" 
WHERE 1=CONTAINS(POINT('ICRS',
 "I/345/gaia2".ra,"I/345/gaia2".dec), 
 CIRCLE('ICRS', 266.6, 5.7, 3.))
\end{verbatim}

\subsection{PANSTARRS}
\label{app:query_PANSTARRS}
\begin{verbatim}
SELECT 
 "II/349/ps1".RAJ2000, 
 "II/349/ps1".DEJ2000, 
 "II/349/ps1".Qual, 
 "II/349/ps1".e_RAJ2000, 
 "II/349/ps1".e_DEJ2000, 
 "II/349/ps1".Epoch, 
 "II/349/ps1".Ns, 
 "II/349/ps1".Nd, 
 "II/349/ps1".gmag, 
 "II/349/ps1".e_gmag, 
 "II/349/ps1".gKmag, 
 "II/349/ps1".e_gKmag, 
 "II/349/ps1".gFlags, 
 "II/349/ps1".rmag, 
 "II/349/ps1".e_rmag, 
 "II/349/ps1".rKmag, 
 "II/349/ps1".e_rKmag, 
 "II/349/ps1".rFlags, 
 "II/349/ps1".imag, 
 "II/349/ps1".e_imag, 
 "II/349/ps1".iKmag, 
 "II/349/ps1".e_iKmag, 
 "II/349/ps1".iFlags, 
 "II/349/ps1".zmag, 
 "II/349/ps1".e_zmag, 
 "II/349/ps1".zKmag, 
 "II/349/ps1".e_zKmag, 
 "II/349/ps1".zFlags, 
 "II/349/ps1".ymag, 
 "II/349/ps1".e_ymag, 
 "II/349/ps1".yKmag, 
 "II/349/ps1".e_yKmag, 
 "II/349/ps1".yFlags 
FROM "II/349/ps1" 
WHERE (("II/349/ps1".RAJ2000>=264.8) 
 AND ("II/349/ps1".RAJ2000 <= 269.8) 
 AND ("II/349/ps1".DEJ2000>=3.14) 
 AND ("II/349/ps1".DEJ2000<=7.4))
\end{verbatim}

\subsection{2MASS}
\label{app:query_2MASS}
\begin{verbatim}
SELECT 
 "II/246/out".RAJ2000, 
 "II/246/out".DEJ2000, 
 "II/246/out"."2MASS", 
 "II/246/out".Jmag, 
 "II/246/out".e_Jmag, 
 "II/246/out".Hmag, 
 "II/246/out".e_Hmag, 
 "II/246/out".Kmag, 
 "II/246/out".e_Kmag,
 "II/246/out".Qflg 
FROM "II/246/out" 
WHERE 1=CONTAINS(POINT('ICRS',
 "II/246/out".RAJ2000,"II/246/out".DEJ2000), 
CIRCLE('ICRS', 266.6, 5.7, 3.))
\end{verbatim}

\subsection{WISE}
\label{app:query_WISE}
\begin{verbatim}
SELECT 
 "II/328/allwise".RAJ2000,  
 "II/328/allwise".DEJ2000,  
 "II/328/allwise".W1mag,  
 "II/328/allwise".e_W1mag,  
 "II/328/allwise".W2mag,  
 "II/328/allwise".e_W2mag, 
 "II/328/allwise".W3mag,  
 "II/328/allwise".e_W3mag,  
 "II/328/allwise".W4mag,  
 "II/328/allwise".e_W4mag,   
 "II/328/allwise".ccf, 
 "II/328/allwise".ex,  
 "II/328/allwise".var,  
 "II/328/allwise".pmRA, 
 "II/328/allwise".e_pmRA, 
 "II/328/allwise".pmDE, 
 "II/328/allwise".e_pmDE,  
 "II/328/allwise".qph
FROM "II/328/allwise" 
WHERE (("II/328/allwise".RAJ2000>=264.8) 
 AND ("II/328/allwise".RAJ2000 <= 269.8) 
 AND ("II/328/allwise".DEJ2000>=3.14) 
 AND ("II/328/allwise".DEJ2000<=7.4))
\end{verbatim}

\section{Tables}

\subsection{Empirical isochrones}

\begin{table}
\centering
\begin{tabular}{|c|c|c|}
\hline
\hline
 $G$   & $G_{BP}$ & $G_{RP}$\\
 
  [mag] & [mag] &  [mag]\\
\hline
\hline
   7.40 &  7.41 &  7.42\\
   7.87 &  7.89 &  7.86\\
   8.40 &  8.44 &  8.35\\
   8.93 &  9.00 &  8.83\\
   9.46 &  9.55 &  9.31\\
   9.99 & 10.11 &  9.80\\
  10.52 & 10.68 & 10.27\\
  11.05 & 11.26 & 10.73\\
  11.58 & 11.85 & 11.17\\
  12.11 & 12.43 & 11.63\\
  12.63 & 13.01 & 12.09\\
  13.15 & 13.59 & 12.56\\
  13.68 & 14.17 & 13.03\\
  14.19 & 14.78 & 13.46\\
  14.70 & 15.39 & 13.89\\
  15.21 & 16.01 & 14.33\\
  15.71 & 16.64 & 14.75\\
  16.20 & 17.28 & 15.16\\
  16.69 & 17.92 & 15.58\\
  17.18 & 18.56 & 16.01\\
  17.66 & 19.20 & 16.44\\
  18.15 & 19.83 & 16.87\\
  18.66 & 20.42 & 17.34\\
  19.18 & 20.98 & 17.82\\
  19.71 & 21.55 & 18.30\\
\hline
\hline\end{tabular}
\caption{Empirical isochrones of IC~4665 for the photometric bands of the \textit{Gaia} DR2 catalog.}
\label{tab:isoc_empiric_GDR2}
\end{table}
\begin{table}
\centering
\begin{tabular}{|c|c|c|c|c|c|}
\hline
\hline
$i$ & $z$ & $y$ & $J$ & $H$ & $Ks$\\

  [mag] & [mag] &  [mag] &   [mag] & [mag] &  [mag]\\
\hline
\hline
  12.57 & 12.56 & 12.55 & 11.60 & 11.20 & 11.10\\
  13.03 & 12.98 & 12.92 & 11.92 & 11.47 & 11.37\\
  13.49 & 13.38 & 13.30 & 12.25 & 11.73 & 11.65\\
  13.97 & 13.79 & 13.65 & 12.55 & 11.96 & 11.86\\
  14.47 & 14.20 & 14.02 & 12.86 & 12.23 & 12.07\\
  14.98 & 14.64 & 14.42 & 13.19 & 12.51 & 12.33\\
  15.46 & 15.05 & 14.79 & 13.55 & 12.82 & 12.64\\
  15.94 & 15.45 & 15.14 & 13.88 & 13.16 & 12.98\\
  16.40 & 15.85 & 15.50 & 14.20 & 13.53 & 13.32\\
  16.85 & 16.22 & 15.88 & 14.54 & 13.90 & 13.67\\
  17.28 & 16.61 & 16.27 & 14.89 & 14.26 & 14.02\\
  17.73 & 17.00 & 16.64 & 15.24 & 14.61 & 14.36\\
  18.19 & 17.41 & 17.03 & 15.58 & 14.95 & 14.68\\
  18.64 & 17.81 & 17.40 & 15.91 & 15.32 & 15.01\\
  19.09 & 18.19 & 17.78 & 16.26 & 15.66 & 15.36\\
  19.59 & 18.60 & 18.16 & 16.60 & 15.97 & 15.67\\
  20.08 & 19.03 & 18.52 & 16.92 & 16.31 & 15.96\\
  20.55 & 19.42 & 18.90 & 17.27 & 16.65 & 16.26\\
  21.05 & 19.83 & 19.29 & 17.62 & 16.94 & 16.57\\
  21.59 & 20.27 & 19.66 & 17.92 & 17.22 & 16.84\\
  22.16 & 20.70 & 20.10 & 18.19 & 17.51 & 17.08\\
  22.74 & 21.18 & 20.50 & 18.48 & 17.84 & 17.30\\
  23.26 & 21.58 & 20.90 & 18.79 & 18.19 & 17.52\\
  23.77 & 22.00 & 21.30 & 19.08 & 18.48 & 17.76\\
  24.29 & 22.42 & 21.70 & 19.38 & 18.75 & 18.03\\
\hline
\hline\end{tabular}
\caption{Empirical isochrones of IC~4665 for the photometric bands of DANCe (pan-STARRS and 2MASS).}
\label{tab:isoc_empiric_DANCe}
\end{table}

\subsection{Magnirude distribution and mass function}

\begin{table}
\centering
\begin{tabular}{|c|c|c|}
\hline
\hline
  $G$ mag &Density &$\sigma_{Density}$ \\
\hline
\hline
   7.08 & 0.0079 & 0.0024\\
   7.65 & 0.0110 & 0.0029\\
   8.22 & 0.0128 & 0.0031\\
   8.78 & 0.0139 & 0.0033\\
   9.35 & 0.0157 & 0.0033\\
   9.92 & 0.0184 & 0.0035\\
  10.49 & 0.0215 & 0.0040\\
  11.06 & 0.0252 & 0.0043\\
  11.62 & 0.0299 & 0.0043\\
  12.19 & 0.0356 & 0.0043\\
  12.76 & 0.0407 & 0.0048\\
  13.33 & 0.0431 & 0.0056\\
  13.90 & 0.0417 & 0.0057\\
  14.46 & 0.0399 & 0.0052\\
  15.03 & 0.0440 & 0.0048\\
  15.60 & 0.0589 & 0.0049\\
  16.17 & 0.0878 & 0.0057\\
  16.73 & 0.1309 & 0.0069\\
  17.30 & 0.1816 & 0.0088\\
  17.87 & 0.2229 & 0.0109\\
  18.44 & 0.2319 & 0.0116\\
  19.01 & 0.1980 & 0.0102\\
  19.57 & 0.1344 & 0.0077\\
  20.14 & 0.0701 & 0.0054\\
  20.71 & 0.0271 & 0.0033\\
\hline
\hline\end{tabular}
\caption{Apparent $G$ magnitude distribution of IC~4665 (normalised).}
\label{tab:G_dist}
\end{table}
\begin{table}
\centering
\begin{tabular}{|c|c|c|}
\hline
\hline
$i$ mag & Density & $\sigma_{Density}$ \\
\hline
\hline
  12.00 & 0.0017 & 0.0010\\
  12.50 & 0.0044 & 0.0017\\
  13.01 & 0.0088 & 0.0026\\
  13.51 & 0.0140 & 0.0032\\
  14.01 & 0.0182 & 0.0034\\
  14.51 & 0.0278 & 0.0043\\
  15.02 & 0.0518 & 0.0063\\
  15.52 & 0.0873 & 0.0069\\
  16.02 & 0.1282 & 0.0078\\
  16.52 & 0.1741 & 0.0099\\
  17.03 & 0.2225 & 0.0106\\
  17.53 & 0.2527 & 0.0115\\
  18.03 & 0.2419 & 0.0111\\
  18.54 & 0.2062 & 0.0101\\
  19.04 & 0.1691 & 0.0098\\
  19.54 & 0.1272 & 0.0084\\
  20.04 & 0.0863 & 0.0074\\
  20.55 & 0.0548 & 0.0058\\
  21.05 & 0.0355 & 0.0049\\
  21.55 & 0.0262 & 0.0049\\
  22.06 & 0.0192 & 0.0039\\
  22.56 & 0.0126 & 0.0029\\
  23.06 & 0.0070 & 0.0020\\
  23.56 & 0.0037 & 0.0015\\
  24.07 & 0.0023 & 0.0012\\
\hline
\hline\end{tabular}
\caption{Apparent $i$ magnitude distribution of IC~4665 (normalised).}
\label{tab:i_dist}
\end{table}
\begin{table}
\centering
\begin{tabular}{|c|c|c|}
\hline
\hline
  $\log_{10}m$ & Density & $\sigma_{Density}$ \\
\hline
\hline
  -2.00 & 0.0026 & 0.0022\\
  -1.88 & 0.0055 & 0.0040\\
  -1.76 & 0.0086 & 0.0038\\
  -1.64 & 0.0256 & 0.0059\\
  -1.51 & 0.0738 & 0.0118\\
  -1.38 & 0.1328 & 0.0154\\
  -1.27 & 0.1982 & 0.0173\\
  -1.14 & 0.3204 & 0.0208\\
  -1.01 & 0.4604 & 0.0254\\
  -0.89 & 0.5434 & 0.0278\\
  -0.76 & 0.6372 & 0.0289\\
  -0.63 & 0.7113 & 0.0287\\
  -0.51 & 0.7029 & 0.0399\\
  -0.40 & 0.6680 & 0.0367\\
  -0.29 & 0.6258 & 0.0356\\
  -0.17 & 0.5704 & 0.0403\\
  -0.06 & 0.4742 & 0.0401\\
   0.05 & 0.3432 & 0.0333\\
   0.17 & 0.2055 & 0.0266\\
   0.29 & 0.1220 & 0.0210\\
   0.39 & 0.0730 & 0.0145\\
   0.51 & 0.0502 & 0.0119\\
   0.63 & 0.0526 & 0.0131\\
   0.73 & 0.0420 & 0.0111\\
   0.85 & 0.0203 & 0.0064\\
\hline
\hline\end{tabular}
\caption{Present-day system mass function of IC~4665 (normalised).}
\label{tab:mass_function}
\end{table}

\end{appendix}

\bibliographystyle{aa} 
\bibliography{mybiblio.bib}

\end{document}